\documentclass[preprint,superscriptaddress,nofootinbib,preprintnumbers,amsmath,amssymb]{revtex4-1}

\usepackage{amsfonts}
\usepackage{mathrsfs}
\usepackage{leftidx}
\usepackage{amssymb}
\usepackage{placeins}
\usepackage{relsize}
\usepackage{slashed}
\usepackage{multirow}
\usepackage[dvipsnames]{xcolor}

\usepackage[colorlinks]{hyperref}
\usepackage{epsfig}
\usepackage{graphicx}               
\usepackage{url}
\usepackage{float}

\newcommand{\be}{\begin{equation}}
\newcommand{\ee}{\end{equation}}
\newcommand{\bea}{\begin{eqnarray}}
\newcommand{\eea}{\end{eqnarray}}
\newcommand{\beq}{\begin{eqnarray}}
\newcommand{\eeq}{\end{eqnarray}}
\def\bit{\begin{itemize}}
\def\eit{\end{itemize}}
\def\ben{\begin{enumerate}}
\def\een{\end{enumerate}}
\def\trans{{\cal{T}}}

\newcommand{\Eq}[1]{Eq.~(\ref{#1})}

\newcommand\DN[1][\relax]{%
\ifx\relax#1\relax\else{}^{#1}\fi \!X}

\newcommand{\Tbbd}{T_\text{syn}}
\newcommand{\TXbbd}{{T_{X}^{\text{syn}}}}
\newcommand{\Tgambbd}{T_{\gamma}^{\text{syn}}}

\newcommand{\BE}{\textrm{BE}}

\begin{document}


\title{Making Asymmetric Dark Matter Gold:\\
Early Universe Synthesis of Nuggets}
\author{Moira I. Gresham}
\affiliation{Whitman College, Walla Walla, WA 99362}
\author{Hou Keong Lou}
\affiliation{Theoretical Physics Group, Lawrence Berkeley National Laboratory, Berkeley, CA 94720}
\affiliation{Berkeley Center for Theoretical Physics, University of California, Berkeley, CA 94720}
\author{Kathryn M. Zurek}
\affiliation{Theoretical Physics Group, Lawrence Berkeley National Laboratory, Berkeley, CA 94720}
\affiliation{Berkeley Center for Theoretical Physics, University of California, Berkeley, CA 94720}

\begin{abstract} 

We compute the mass function of bound states of Asymmetric Dark Matter---nuggets---synthesized in the early Universe.  We apply our results for the nugget density and binding energy computed from a nuclear model to obtain analytic estimates of the typical nugget size exiting synthesis.  We numerically solve the Boltzmann equation for synthesis including two-to-two fusion reactions, estimating the impact of bottlenecks on the mass function exiting synthesis. 
These results provide the basis for studying the late Universe cosmology of nuggets in a future companion paper. 

\end{abstract}

\maketitle
\tableofcontents

\section{Introduction}

Searches for dark matter (DM)---whether through DM annihilations to Standard Model (SM) particles in our galaxy, through observations of structure, through collider searches, or through DM interactions with SM nuclei in underground detectors---have focused on DM that behaves as a single massive particle or as a coherent field configuration.  The weakly interacting massive particle (WIMP) and the axion have served as primary motivators of these search techniques; in addition to being well-motivated candidates, they provide sharp predictions for experiments probing the nature of dark matter.  It is, however, important to explore well-motivated new ideas for dark matter candidates that lead to qualitatively different experimental signatures of dark matter, and inform new search strategies.

Even modest changes in the nature of the dark sector can have radical implications for cosmology and its associated observational constraints.  In this paper, we consider a DM particle with a particle-anti-particle asymmetry (as in Asymmetric Dark Matter (ADM), see {\em e.g.} \cite{Petraki:2013wwa,Zurek:2013wia} for reviews) that self interacts through an attractive force.  Similar to SM nuclei, given a sufficiently strong attractive force, DM particles bind together to form composite states.  The size of these composite states depends both on  the strength and range of the force, and on the presence or absence of bottlenecks.  For example, in the SM, only small nuclei are synthesized in the early Universe because the $A=5,~8$ nuclei are unstable; three helium nuclei must fuse to carbon in order to surpass this bottleneck.  In a dark sector, however, such bottlenecks may not be present.  This is especially true if a repulsive long-range force like electromagnetism is absent, and if the force mediator binding the dark nuclei is longer range than the nuclear force mediated by QCD mesons. 

We study the synthesis of the dark nuclei, which, as in Ref.~\cite{Wise:2014jva}, we refer to as ``nuggets.''  Our goal is to obtain results that connect the size of the synthesized nuggets to a simple UV complete model; we employ a model with fermionic ADM and a scalar mediator with arbitrary mass that mediates an attractive dark force.  We draw on the results of our companion paper  \cite{Gresham:2017zqi}, which computes the relevant nugget properties---notably the saturation density and binding energy---using the $\sigma$-$\omega$ model from nuclear physics.  The $\sigma$-$\omega$ model, named for the lightest scalar and vector QCD mesons mediating the most important attractive and repulsive interactions that bind large nuclei, employs Relativistic Mean Field Theory (RMFT) in order to solve the equations of motion that determine bulk properties of ground states of nuclei. 
We use results from \cite{Gresham:2017zqi} along with the Compound Nucleus Model to obtain approximations for the fusion and splitting cross-sections.

Synthesis becomes efficient once the temperature of the Universe drops below the binding energy of the nugget.  Utilizing simple analytic arguments, along with our results in Ref.~\cite{Gresham:2017zqi}, we are able to obtain an estimate for the typical nugget size just before the era of structure formation.  We also obtain a distribution of nugget sizes by solving the Boltzmann equations for synthesis numerically.  Using the results of this simulation and argument by analogy to SM nucleosynthesis, in our Coulomb repulsion-free scenario, we argue that a substantial bottleneck is likely to occur only if multiple states adjacent in size ({\em e.g.}~$N=3,4$) are unstable. Superficial bottlenecks, corresponding to isolated states being hard to form (either due to a very low formation cross section or absence of a stable state), do not significantly affect the distribution of nugget size (or mass function) at the end of synthesis.

Our main results are shown in Figs.~\ref{fig: Nmax} and \ref{fig: bottleneck plot}, for the cases of a presence or absence of a bottleneck to synthesis at low $N$.  In parameter ranges where two-particle bound states are easily formed in the early universe (see \cite{Wise:2014jva}), assuming that other substantial bottlenecks at low $N$ do not occur, large nuggets can be synthesized in the early universe in the presence of sufficiently attractive and low mass forces. For example, as shown in Fig.~\ref{fig: Nmax}, assuming DM mass $m_X\sim$ GeV, DM-mediator fine structure coupling constant $\alpha_\phi \sim 0.3$, and mediator mass $m_\phi \sim $ MeV, the typical nugget size exiting early Universe synthesis is order $N \sim 10^{12}$ or $M_N \sim 10^{11}$ GeV, with the sizes scaling as $N\sim(\BE_2^{3/2}/\bar m_X^{7/2})^{6/5}$ and $M_N \sim \bar m_X N$, where the two-particle bound state binding energy that sets the synthesis temperature is $\BE_2 \sim \alpha_\phi^2 m_X/4$ and the nugget saturation mass scale is $\bar m_X \sim \alpha_\phi^{-1/4} \sqrt{m_X m_\phi}$.  Since large nugget binding energies are typically quite large when the DM/mediator mass hierarchy is large, the nugget masses can be significantly smaller than $m_X N$. 

If synthesis proceeds at all, we find that nuggets are generically synthesized past their saturation size, where the force range, $m_\phi^{-1}$, sets the nugget size and binding energy along with the coupling and DM mass. 
Nuggets bound by a scalar mediator much lighter than the dark matter constituent saturate at a larger $N$, but by virtue of the effectively stronger coupling in this limit, synthesis to large nuggets occurs more readily.   Simulating synthesis with fusion through both mediator and small nugget emission, we show that the spectrum of nugget sizes exiting synthesis can be fairly broad. The mass function exiting synthesis is shown in Fig.~\ref{fig:mass_function}. This may have implications for the late Universe cosmology and detection prospects, which we study in a future companion paper \cite{lateuniverse}.  Substantial bottlenecks at low $N$, on the other hand, can lead to a bimodal distribution of dark matter exiting early universe synthesis, with the majority of DM residing in the form of small-$N$ nuggets and a subdominant population of even larger nuggets---for example, as shown in Fig.~\ref{fig: bottleneck plot}, with typical size $N \sim 10^{20}$ or $M_N \sim 10^{19}$ GeV for the benchmark quoted above.  

Dark matter nugget properties and synthesis (``darkleosynthesis'') have been explored elsewhere, though with more limited assumptions for obtaining quantitative results.  Early universe synthesis of two-particle bound states was addressed in \cite{Wise:2014jva}; we will utilize these results in the initial step of our analysis.  Ref.~\cite{Wise:2014ola} considered larger $N$ bound states, but did not address the saturation limit.  We will find, however, that nugget synthesis typically terminates with saturated nuggets, even when the attractive force is very light relative to the DM.  Ref.~\cite{Hardy:2014mqa} considered the synthesis of nucleus-like nuggets, assuming dark nuclei properties (notably the saturation density and binding energies) that directly mirror SM nuclei.  We utilize similar analysis techniques, but obtain different analytic and numerical results in two primary ways.  First,  and most importantly, we apply the results of \cite{Gresham:2017zqi} for the nugget properties, which implies larger synthesized nugget sizes (by several orders of magnitude).  Second, instead of employing a simple fusion model with geometric cross-sections involving a single mediator emission, we consider the Compound Nucleus Model and show that multiple mediator emissions are dominant with possible light dark nugget emission, analogous to nuclear fusion through nucleon or $\alpha$ particle emission.  Ref.~\cite{Krnjaic:2014xza} studied the synthesis of dark nuclei modeled on SM nuclei, with a dark confining force binding the composite dark baryons into nuclei and an additional weak dark electromagnetism enabling ``di-darkleon'' fusion.  Lastly, Ref.~\cite{Detmold:2014qqa} studied the synthesis of dark spin-0 deuterium forming in a two-flavor, two-color, dark QCD.

The outline of this paper is as follows.   In the next section we review the features relevant for synthesis of our nuclear model for ADM nuggets.  Then in Sec.~\ref{sec:cosmology} we outline the conditions for synthesis to begin in the context of our simple UV complete model.  In Sec.~\ref{sec:synthesis}, we set up the Boltzmann equations with the appropriate fusion and dissociation rates before solving them.  We utilize these results to obtain analytic estimates for the typical nugget size.  We conclude with an eye toward future work exploring the impact of dark nuggets on stellar and structure formation.

\section{A Nuclear Model for Asymmetric Dark Nuggets}\label{sec: model}

We will consider a model with a single Dirac fermion with attractive self interactions mediated by a lighter, real scalar, governed by 
\be
  \mathcal{L} = \bar X \left[i\slashed\partial - (m_X -g_\phi \phi)\right]X +
\frac{1}{2}
(\partial \phi)^2 - \frac{1}{2} m_\phi^2\phi^2  - V(\phi).\label{eq:lagrangian}
\ee
As discussed in detail in \cite{Gresham:2017zqi}, large collections of dark matter can form stable bound states when $\alpha_\phi {m_X^2 \over m_\phi^2} \gtrsim 2.6$, where $\alpha_\phi = g_\phi^2/4\pi$. Two-body bound states form when ${\alpha_\phi \over 2}{m_X \over m_\phi}\gtrsim 0.84$ (assuming perturbative coupling; see \cite{Wise:2014jva} and references therein). Here we are interested in the synthesis of such bound states in the early Universe. As we will see, for synthesis to begin with two-particle bound states, it is important that the force mediator be light enough to be produced on shell in the fusion process, such that that $\BE_2 \gtrsim m_\phi$. Since, for perturbative coupling, $\BE_2 \sim \alpha_\phi^2 m_X$, one generally needs $m_X > m_\phi$.   

It is also natural to consider a strongly coupled dual of this model where $X$ is the analog of a nucleon and $\phi$ is the analog of the lightest scalar meson, $\sigma$ (or $f_0$). In a composite model, one expects other meson degrees of freedom such as vectors (like the $\omega$) and pseudoscalars (like pions) to mediate repulsive and/or spin-dependent interactions of comparable importance. If there is an additional approximate symmetry in the dark sector, the analog of isospin-dependent interactions may also be important. One also naively expects a mass hierarchy between the dark matter constituents (baryons) and force mediators (mesons) to be moderate, and for the masses of the mesons to be of very similar size. As discussed in \cite{Gresham:2017zqi}, if this is the case, it is unlikely that there is a viable region of parameter space in which any of the mesons is lighter than the two-body binding energy, which will stifle early Universe synthesis. 

In the SM, deuterium forms through $p+n \rightarrow D + \gamma$. The deuterium binding energy $\sim 2$ MeV is significantly smaller than $m_{\pi^0} \sim 135$ MeV. Without electromagnetism, fusion into deuterium would require $\pi^0$ emission. Such a process would then only be efficient at temperatures near $m_\pi$, where dissociation would dominate.
So, taking a cue from the SM, why not just add a dark photon in order to enable the first step of synthesis? (This is precisely the scenario considered in \cite{Krnjaic:2014xza}.) There are costs. Including a dark photon will destabilize dark nuggets of sufficiently large size---just as electromagnetism helps to destabilize large nuclei. The larger the coupling, the smaller the size at which nuggets will destabilize. So allowing for very large nuggets requires very small coupling. But the smaller the coupling  the smaller the 2-body fusion cross section. 

Due to these complications, we will focus on the case of an attractive force mediator only, where there is a large parameter space for efficient nugget synthesis. We will restrict our attention to perturbative couplings, where we have a good handle on the 2-body bound state properties that govern the initiation of early Universe synthesis, even though the RMFT used to deduce properties of large nuggets is valid also for nonperturbative coupling.  In our synthesis estimates that follow, the behavior of binding energy and fusion cross sections as a function of nugget size will be important. Thus here we first summarize these features of nuggets, justified and presented in more detail in \cite{Gresham:2017zqi}. 

\subsection{Binding Energy and the Liquid Drop Model}

In \cite{Gresham:2017zqi}, relativistic mean field theory (RMFT) was used to derive the behavior of nugget structure (density, size) and binding energy as a function of the nugget number, $N$. RMFT applied to nucleons has been used to accurately model bulk properties of large nuclei such as binding energy, density, and the saturation property of nuclei---that the density is relatively constant as a function of mass number \cite{Negele:1986bp}.  

As one expects since the only force in play is attractive, the binding energy per dark number increases as a function of $N$. At some $N$ determined roughly by $N_\text{sat} \sim \left( \alpha_\phi {m_\phi^2 \over m_X^2} \right)^{- 3/4}$, the binding energy per dark number levels off, asymptoting towards a constant determined as a fraction of $m_X$ by the combination of parameters, $C^2 \equiv \alpha_\phi {m_X^2 \over m_\phi^2} $ (a larger fraction for larger $C^2$). At $N > N_\text{sat}$, the density as a function of $N$ also approaches a constant, exhibiting saturation behavior: adding further constituents does not change the nugget density but simply increases the nugget size as $R(N) \propto N^{1/3}$. For $N>N_\text{sat}$, just as for large nuclei which exhibit saturation, a liquid drop model gives a good description of the nugget binding energy:
\beq
N m_X - \BE_N \approx N \mu_0 + \epsilon_\text{surf} N^{2/3} \label{eq: liquid drop}
\eeq
where $0 < \mu_0 < m_X$ is a bulk energy constant equivalent to the chemical potential of infinitely large nuggets and $\epsilon_\text{surf} N^{2/3} > 0$ (with $N_\text{sat}^{1/3}(m_X - \mu_0) \gg \epsilon_\text{surf}$) is a correction term proportional to the surface area of the nugget  that takes into account  the dearth of close-range attractive interactions of constituents at the surface which would have decreased the energy of the configuration. The infinite matter chemical potential $\mu_0$ approaches $m_X$ as $C^2 \rightarrow \infty$ and the surface energy constant $\epsilon_\text{surf}$ grows with decreasing $m_\phi / m_X$ and decreasing $\alpha_\phi$.  (See Table I of \cite{Gresham:2017zqi}.) 

\subsection{Saturation Properties of Nuggets}\label{sec: saturation properties}

In the following sections, we will show that in the range of parameter space in which 2-body synthesis proceeds easily, nuggets are generically synthesized up to sizes beyond which they exhibit saturation behavior. We are able to self-consistently estimate the results of synthesis in terms of the saturation properties of nuggets: chiefly, in terms of average nugget mass per dark number, $\bar m_X = m_X - \BE_N / N \sim \mu_0$ and nugget length scale $r_0 \approx \left( {4 \over 3} \pi n_\text{sat} \right)^{-1/3}$, with $n_\text{sat}$ the saturation number density. 

In the limit where large nuggets are strongly bound (which is generally true when $\alpha_\phi < 1$ and $\BE_2>m_\phi$), the saturated nugget parameters $r_0$ and $\bar m_{X 0}$ are given approximately by,
\beq
{\bar m_{X 0}} \approx \left({3 \pi \over 2 \alpha_\phi}\right)^{1/4} \sqrt{m_X m_\phi} \qquad \text{and} \qquad r_0  \approx \left(9 \pi \over 4 \right)^{1/3}{1 \over \bar m_{X 0}}. \label{eq: r0 estimate}
\eeq  
The $\bar m_{X 0} \rightarrow m_X$ limit represents the limit of no binding, where the above equations are invalid. 

Inclusion of a quartic potential of the form $V(\phi) = {4 \over 3} \lambda_4 \alpha_\phi^2 \phi^4$  leads to an effective scalar mass, 
\beq
m_{\phi \text{eff}} = \sqrt{m_\phi^2 + \lambda_4 {2  \alpha_\phi \over 3 \pi} (m_X-m_*)^2},
\eeq
where $m_*$ is the effective DM mass, related to the scalar field VEV through $m_* = m_X - g_\phi \langle \phi \rangle$. The equations of motion guarantee that $m_* \geq 0$. In the limit where $\alpha_\phi m_X^2 \gg m_\phi^2$ and $\lambda_4 \ll 1$, $m_*  \ll m_X$ and we find \Eq{eq: r0 estimate} holds if we replace $m_\phi$ with $m_{\phi \text{eff}}$.
See  \cite{Gresham:2017zqi} for details.

The saturation size, $N_\text{sat} \sim (r_0 m_{\phi \text{eff}})^{-3}$, corresponding to the point where nugget radius exceeds the force range, $R \gtrsim m_{\phi \text{eff}}^{-1}$, is also important in checking the consistency of our estimates. In the synthesis estimates that follow, we will assume $\lambda_4 \ll 1$ so that the effects of the quartic potential are essentially encapsulated in $m_{\phi \text{eff}}$. We will frame our analysis as if the potential is completely absent, but we note that our results hold equally well when a quadratic potential with $\lambda_4 \lesssim 0.01$ is included through taking $m_\phi \rightarrow m_{\phi \text{eff}}$. It should also be noted that, even if a large hierarchy between $m_\phi$ and $m_X$ is achieved, the quartic coupling will limit the size of the $m_X/m_{\phi\text{eff}}$ hierarchy relevant to nugget properties.  

\section{Conditions for Synthesis}\label{sec:cosmology}

Here we discuss conditions for initiating synthesis with formation of two-dark-nugget bound states, $\DN[2]$, the temperature at which this initiation occurs, and possible bottlenecks at low dark nugget number. We will follow nuclear physics convention and denote each dark nugget species as $\DN[N]$, where $N$ is the dark nugget number.

\subsection{Conditions for Beginning Synthesis}\label{sec: synthesis start conditions}
Nugget synthesis begins in the early Universe when $\DN[2]$ bound states form. This very first stage of synthesis corresponds to passing the two-DN bottleneck, analogous to the deuterium bottleneck in BBN. 
The bound state $\DN[2]$ can begin to accumulate if the $\DN[2]$ formation rate exceeds the Hubble expansion rate when the dissociation rate drops below the formation rate. This occurs when the number density of mediators energetic enough to dissociate the $\DN[2]$ drops below the $\DN[2]$ equilibrium number density. 

\begin{figure}
\includegraphics[width=0.7\textwidth]{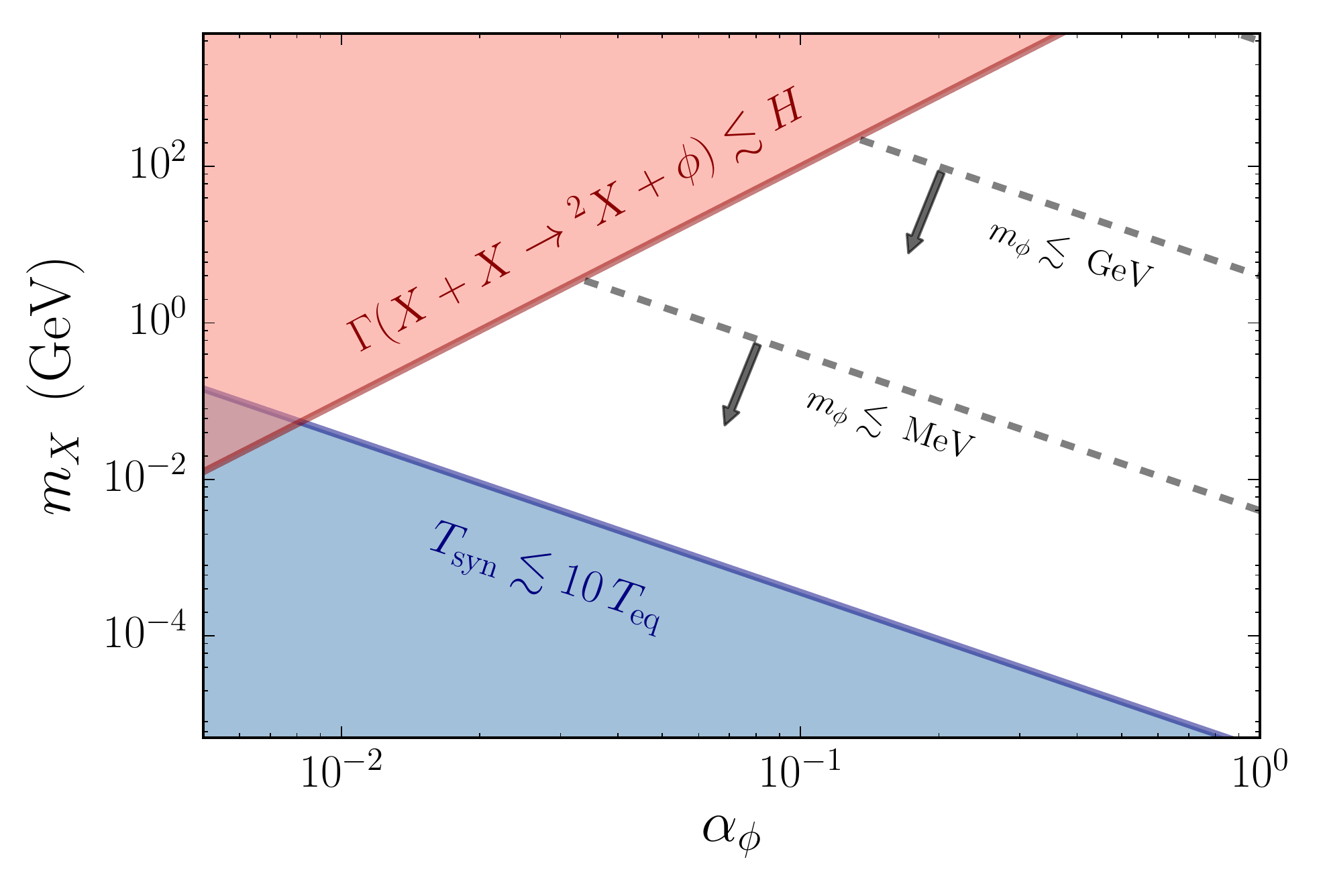}
\caption{Approximate conditions for synthesis to begin with efficient formation of $\DN[2]$. The top line corresponds to the condition for the formation rate to ever be larger than the Hubble rate in the $\BE_2 \gg m_\phi$ limit (\Eq{eq: synthesis condition 2}). The bottom line shows the condition for synthesis to begin when the temperature is an order of magnitude larger than $T_\text{eq}$. The dashed lines correspond to $\BE_2 = m_\phi$; at least roughly speaking, $\BE_2>m_\phi$ (\Eq{eq: synthesis condition 1}) in order for the dissociation rate to ever dip below the formation rate.}\label{fig: synthesis conditions}
\end{figure}

Because formation through the process $X + X \rightarrow \DN[2] + \phi$ is typically inefficient when $m_\phi\gtrsim \BE_2$, as argued in detail in Ref.~\cite{Wise:2014jva}, efficient two-DN synthesis generally requires 
\beq
\BE_2 \approx \alpha_\phi^2 {m_X \over 4} \gtrsim m_\phi, \qquad \text{(first $\DN[2]$ synthesis condition)}
\label{eq: synthesis condition 1}
\eeq
where we have used the expression for 2-body Yukawa bound state binding energy in the Coulomb (hydrogen-like) limit, which is self consistent as long as the coupling is perturbative. In addition, for the formation rate to exceed the Hubble rate at any temperature below the $X$ freeze-out temperature, we also require (see \cite{Wise:2014jva}),
\beq
\alpha_\phi \gtrsim 0.1 \left(m_X \over 100~\text{GeV}\right)^{1/3} \qquad \text{(second $\DN[2]$ synthesis condition)}.
\label{eq: synthesis condition 2}
\eeq  Fig.~\ref{fig: synthesis conditions} shows these $\DN[2]$ synthesis conditions, along with the region of parameter space for which the $\DN[2]$ synthesis temperature---as we discuss in the next section---is an order of magnitude larger than the temperature at matter-radiation equality, $\Tbbd > 10 \; T_\text{eq}$. It is interesting to note that synthesis can begin significantly before the end of radiation domination only if $\alpha_\phi \gtrsim 0.01$.

\subsection{Synthesis Temperature}\label{sec: synthesis temperature}

Since dissociation of $\DN[2]$ must become inefficient relative to formation in order for synthesis to begin, and because dissociation becomes inefficient only once the temperature drops below the binding energy, the big bang darkleosynthesis temperature, $\Tbbd$, is typically set by  
$\BE_2$. This is analogous to BBN, where $T_\text{BBN}$ is set by the deuteron binding energy, $\BE_D \sim 2$ MeV. 

More specifically, one can estimate $T_\text{BBN}$ by setting the number density of photons with energy greater than $\BE_D$ equal to the deuteron equilibrium number density, and solving for the temperature. Because the baryon to photon ratio is so small, the temperature must fall to order $T_{\rm BBN} \sim \BE_D/37$.  Following the same logic, and ignoring all but the $\DN[2]$ ground state, chemical equilibrium implies a Saha relation
\beq
n_{\DN[2]} = \frac{n_{\DN[1]}^2}{4} \left({2  - {\BE_2 \over m_X} }\right)^{3/2} \left({{m_X T_X \over 2 \pi}}\right)^{-3/2}e^{\BE_2/T_X}, \label{eq: saha}
\eeq where we have taken the $\phi$ chemical potential to be zero, which is valid as long as number changing processes for $\phi$ are still efficient. At the very beginning of synthesis the total dark number density $n_X$ is dominated by unbound $X$s so that we may take $n_{\DN[1]} \approx n_X = {\Omega_\text{DM} \over \Omega_B}{m_p \over  \bar m_X} n_B$ where $\bar m_X$ is the average mass per dark number at the end of synthesis.

The equilibrium number density of $\phi$s with enough energy to dissociate $\DN[2]$ is 
\beq
n_\phi(\epsilon > \BE_2) = {1 \over 2 \pi^2} \int_{\BE_2>m_\phi}^\infty {\sqrt{\epsilon^2 - m_\phi^2} \over e^{(\epsilon - \mu_\phi)/T_\phi} - 1} \epsilon \, d \epsilon \approx {e^{-\BE_2 / T_X} ({\BE_2})^2 T_X \over 2 \pi^2}\,,
\eeq where again we have set $\mu_\phi = 0$ and $T_\phi = T_X$. 
Setting $n_\phi(\epsilon > \BE_2) = n_{\DN[2]}$ leads to the condition for synthesis temperature,
\beq
 {e^{-2 \BE_2 / \TXbbd} \left({{\BE_2} \over \TXbbd }\right)^{7/2} } \approx 10^{-15} \left({g_{*S}(\Tgambbd) \over 10.75}\left({\Tgambbd \over \TXbbd}\right)^3 {\text{GeV} \over \bar m_X} \right)^2 \left( \BE_2 \over m_X \right)^{3/2}\,,
\eeq where we have assumed $\BE_2 \ll m_X$. For $\text{MeV} \lesssim m_X \lesssim \text{TeV}$, $0.01 \lesssim \alpha_\phi \lesssim 0.3$, and including other constraints for $\DN[2]$ synthesis discussed above (Eqs.~\eqref{eq: synthesis condition 1} and \eqref{eq: synthesis condition 2}), we find that $ \BE_2 / \TXbbd \sim 10$ to $30$.

Here we briefly discuss the relationship between the DM and SM bath temperatures. Suppose the dark sector kinetically decouples from the SM at temperature $T_d$, when the DM bath is still relativistic. Then, while the DM bath is still relativistic, the ratio of dark to SM temperatures changes only when some relativistic species falls out of chemical equilibrium and dumps its entropy in one bath or the other. Supposing only the SM bath is heated relative to the DM bath, for example, then ${T_\gamma \over T_X} \sim \left(\frac{g_{\rm SM}(T_d)}{g_{\rm SM}(T_\gamma)}\right)^{1/3}$ where $g_{\rm SM}(T_d)$ counts the number of relativistic degrees of freedom in the SM when the two sectors decoupled and $g_{\rm SM}(T_\gamma)$ counts them at temperature $T_\gamma$. When $\phi$ becomes nonrelativistic and decays, its entropy is dumped either into some hidden sector bath (which would heat that bath relative to the SM bath) or into the SM bath, heating it further.  If $m_\phi \gtrsim 5 \mbox{ MeV}$ when it decays to the SM (whether electrons or neutrinos), there is little effect on BBN or CMB measurements of the number of relativistic degrees of freedom, $N_{\rm eff}$ \cite{Boehm:2013jpa}; if, on the other hand, $m_\phi \lesssim 5 \mbox{ MeV}$ and the decay is to neutrinos or to additional radiation in the dark sector, there may be observable signals in next generation CMB experiments, depending on the decoupling temperature of the dark sector from the SM. In this paper we take an agnostic view towards the exact dynamics that determine the ratio of temperatures, but note that in many plausible scenarios, ${T_\gamma \over T_X} \sim \text{few}$, and we take ${T_\gamma \over T_X} \approx 1$ in several illustrative plots that follow; this ratio has little quantitative or qualitative impact on our results.

Once $T \lesssim \TXbbd$, larger $N$ nuggets will efficiently form unless there is an additional bottleneck at low $N$ analogous to the $A\sim 5$ bottleneck for BBN.   This is because, as shown in \cite{Gresham:2017zqi}, the binding energy per particle of larger nuggets increases monotonically as a function of $N$, asymptoting to the saturation value at very large $N$.   Up to angular momentum-dependent effects, it is thus energetically favorable to form successively larger nuggets, with the threshold for dissociation of nuggets growing ever larger than the temperature.\footnote{This is in contrast to large nuclei in the SM, where binding energy per nucleon grows with mass number $A$ up until a global maximum near $A \sim 60$; the drop is due to Coulomb (repulsion) energy and symmetry energy associated with neutron-proton imbalance.}  By arguing through analogy with the SM, in the next subsection we discuss the circumstances under which a nugget bottleneck could appear.

A feature of large nuggets discussed in \cite{Gresham:2017zqi} is that the total binding energy per dark (baryon) number can be a substantial fraction of $m_X$. Thus if very large nuggets are synthesized, because all of this binding energy is released in the synthesis process, one might worry that substantial heating of the dark bath relative to the SM bath could occur. However, under reasonable assumptions, we will see that the binding energy release results in at most an $\mathcal{O}(1)$ change in the dark sector temperature. Assuming that the dark bath consists of only the scalar mediators, $\phi$, and the dark matter, we may place an upper bound on the change in temperature:
\beq
\Delta T_X \lesssim {n_X \over n_\phi + n_X}\bigg|_{\TXbbd + \Delta T_X} {\BE_{N\,\text{sat}} \over N}\,, \label{eq: frac temp change}
\eeq where ${\BE_{N\,\text{sat}} / N}$ is the saturation binding energy per dark number.\footnote{If all of the dark matter were bound in nuggets of effectively infinite dark number and there were no additional components of the dark bath, then the inequality would be saturated.} For ${n_\phi \over n_X} \gg {{\BE_{N\,\text{sat}} / N} \over \TXbbd}$, the temperature change will be negligible. The analogous statement for nucleosynthesis is that ${n_\gamma \over n_B} \gg {\BE_{{}^4\text{He}} /4 \over T_\text{BBN}} \sim 100$, which, of course, is easily satisfied. Here since $\phi$ is massive and because the binding energy per dark number can be an order one fraction of $m_X$, we can have ${\BE_{N \, \text{sat}}/N \over \TXbbd} \sim {m_X \over \BE_2/30} \sim {100 \over \alpha_\phi^2} $,\footnote{Though recall  $\alpha_\phi \gtrsim 0.01$. See Fig.~\ref{fig: synthesis conditions}.}  so the condition ${n_\phi \over n_X} \gg {\BE_{N \, \text{sat}}/N  \over \TXbbd}$ does not necessarily  hold. However, since the asymmetric abundance $n_X$ has frozen out well before synthesis begins,\footnote{Since $\Tbbd \lesssim \BE_2/15 \sim \alpha_\phi^2 m_X/60$, $X$ is nonrelativistic at the onset of synthesis when the coupling is perturbative. During synthesis, the average mass per dark number may decrease as more $X$ are bound into (more and more tightly bound) nuggets, but any individual DM nugget will always have mass greater than or equal to $m_X$.} heat release increases the $\phi$ number density but not the $X$ number density, which serves to limit the overall temperature change. Taking the equilibrium number density $n_\phi(T) = {1 \over 2 \pi^2} \int_{m_\phi}^{\infty} {\sqrt{\epsilon^2 - m_\phi^2} \over e^{\epsilon / T} -1} \epsilon \, d \epsilon$ and $n_X = {\Omega_\text{DM} \over \Omega_B}{m_p \over m_X } n_B\big|_{T_\gamma = \TXbbd}$, we find that $\Delta T_X \lesssim \TXbbd$ in the parameter range of interest. The temperature change can approach order 1 when $\phi$ is nonrelativistic at the onset of synthesis (when $\BE_2 \sim m_\phi$, corresponding to $m_\phi/m_X$ at the upper range of what we consider) or when $\phi$ is very highly relativistic $m_\phi \lll m_X$ at the onset of synthesis. In the relativistic $\phi$ limit, 
\beq
{n_X (\TXbbd) \over n_\phi(\TXbbd + \Delta T_X) } \sim {\eta { \Omega_\text{DM} \over \Omega_{B}}{m_p \over \bar m_X  } { g_{*S}(\Tgambbd)}\left({\Tgambbd \over \TXbbd} \right)^3 \left(1 + {\Delta T_X \over \TXbbd} \right)^{-3}} 
\eeq
where $\bar m_X$ is the average mass per dark number and $\eta$ is the baryon-to-photon ratio. When $m_\phi \lll m_X$, then $\bar m_X = m_X - \BE_{N \text{sat}}/N \sim \alpha_\phi^{-1/4} \sqrt{m_\phi m_X}$ (see \Eq{eq: r0 estimate}) assuming most DM is bound into saturated nuggets at the end of synthesis. Though $n_X/n_\phi$ can be larger than naively expected, as long as $\sqrt{m_X m_\phi}  \gg 10^{-10} m_p \left({\Tgambbd \over \TXbbd} \right)^3$, then $ n_X \ll  n_\phi$.

Putting this all together in the $m_\phi \lll m_X$ limit we then have
\be
 {\Delta T_X \over \TXbbd} \left(1 + {\Delta T_X \over \TXbbd} \right)^3 \sim 10^{-9} {g_{*S}(\Tgambbd)}\left({\Tgambbd \over \TXbbd} \right)^3 {m_p \over \sqrt{m_X m_\phi}}\alpha_\phi^{1/4} {\BE_{N \text{sat}}/N \over \TXbbd} \qquad (m_\phi \lll m_X).
\ee
As discussed above, ${\BE_{N \text{sat}} / N \over \TXbbd} \rightarrow {m_X \over  \TXbbd} \sim 100/\alpha_\phi^2$ in the $m_\phi \lll m_X$ limit. So {\em e.g.} for $\alpha_\phi \sim 0.1$, one could approach an order one temperature change when $\sqrt{m_\phi m_X} \lesssim 10^{-4}$ GeV or an order of magnitude when $\sqrt{m_\phi m_X} \lesssim 10^{-8}$ GeV.  It is important to note that the heat per dark number, $\sim \BE_{N \, \text{sat}}/N $,  is not released all at once, but rather in increments over the entire synthesis process up to large nuggets.  Since on average the dissociation temperature rises as a function of nugget size, the heat release should not delay synthesis.

\subsection{Bottlenecks at Low Dark Nugget Number?}

\def\he{{}^4\text{He}}
\def\hethree{{}^3\text{He}}
\def\hefive{{}^5\text{He}}

By arguing through analogy with the SM, here we discuss the circumstances under which a dark nugget bottleneck could appear.
First we briefly review the nature of the SM BBN bottlenecks.
BBN proceeds when the temperature drops below the deuteron binding energy (order MeV) so that deuteron dissociation is suppressed. Once a substantial fraction of deuterons exist, almost all free neutrons are incorporated into $\he$ (with intermediate steps of  {\em e.g.}~D(D,n)$\hethree$, $\hethree$(D,H)$\he$). 
A few factors prevent any significant amount of synthesis to higher-$A$ isotopes.
First, the Coulomb barrier 
is substantial at $T_\text{BBN}$ and becomes rapidly more substantial as $A$ and $Z$ grow. 
Second, building up from $\he$ to larger nuclei through Coulomb-barrier-free neutron capture is stifled because $\hefive$, with its unpaired neutron, is unstable to decay to $\he + n$.  Additionally, a $\he +\he$ fusion process is endothermic, 
including ${}^8\text{Be} + \gamma$, such that ${}^8\text{Be}$ is unstable. 
Taken together, these facts imply the presence of bottlenecks not only at $A = 2$, but also at $A = 5$ and $A = 8$.  This guarantees a small abundance of nuclei larger than $\he$ exiting BBN.

We can consider  the existence of bottlenecks in nugget synthesis by analogy with the SM. By construction there is no obstruction due to a Coulomb barrier.  However, fluctuations in binding energy as a function of nugget size, leading to unstable states analogous to $A=5,8$ for nuclear matter, could lead to substantial bottlenecks. The absence of stable $A=5,8$ nuclei can be understood qualitatively within the shell model of the nucleus. Within the shell model, constituents are treated as non-interacting, but each is under the influence of an emergent potential. Constituents fill up available states from low to high energy, obeying Pauli's exclusion principle. The existence of many approximately degenerate energy eigenstates results in some states having significantly larger binding energy per constituent than their neighbors with slightly larger nugget number. The constituent number $N_i$ of these exceptional states are referred to as magic numbers. For spherically symmetric bound states, the first two magic numbers are always $2$ and $8$ (corresponding to doubly magic ${}^4\text{He}$ and ${}^{16}\text{O}$ nuclei), corresponding to the filling of $l=0$ and $1$ states. For larger $N$, the locations for the magic numbers depend on the potential and spin-orbit coupling, and may scale like $N_i \sim i^2$ to $i^3$ (for Coulomb and quadratic potential). 

States with size just over or equal to a sum of magic numbers are prone to  instability or metastability. Isospin slightly complicates the story for nuclei, but consider states closely related to doubly-magic ${}^4\text{He}$. The doubled state, ${}^8\text{Be}$, decays to $2 {}^4\text{He}$. States with one extra neutron or proton (${}^5\text{He}$ or ${}^5\text{Li}$) decay to ${}^4\text{He}$ plus a free nucleon. For the $A=5$ states, pairing also plays a role; nuclei with unpaired nucleons of given isospin are less strongly bound. This is because (pseudoscalar) pions mediate an attractive interaction between opposite-spin nucleons, effectively reducing the energy of the configuration. Since we have only a scalar mediator, it is unclear whether an analog of a pairing effect might exist.\footnote{By virtue of paired constituents being in close proximity, since real scalars will  mediate a strong attractive interaction independent of spin, one can imagine that a similar pairing effect might arise.} In any case, in the context of small nuggets, if the binding of $\DN[4]$ is smaller than twice the binding energy of $\DN[2]$, the decay process $\DN[4] \rightarrow \DN[2] + \DN[2]$ will become possible. Similarly, perhaps the rest energy of $\DN[3]$ is greater than that of $\DN[2]+X$ since the third nucleon in $\DN[3]$ sits alone, unpaired, in an $l=1$ shell. For larger $N$, the degeneracy of energy eigenstates is typically broken more strongly and the spacing of energies becomes smaller so that magic numbers become rare and unimportant.

Since the system under consideration contains only one type of constituent (isospin zero) instead of two as in the SM, as just discussed,  the analogue to unstable $A=5$ and $A=8$ states for nuclei is unstable $N=3$ and $N=4$ nuggets. 
 If both ${}^3X$ and ${}^4X$   are unstable and have small lifetimes in comparison to the cross section for fusions $\left( n_{3(4)} v \sigma(\DN[3(4)]+\DN[2] \rightarrow \DN[5(6)]+\DN[0]) \right)^{-1}$, then there is a bottleneck to fusion of larger nuggets.  In the absence of a Coulomb barrier, if only \emph{one} of $\DN[3]$ and $\DN[4]$ were unstable, it would be easier to synthesize past this artificial bottleneck (e.g.~$\DN[3]+\DN[2] \rightarrow \DN[5]+\phi$ or $\DN[2]+\DN[2] \rightarrow \DN[4]+\phi$).\footnote{Though note that rates for these processes could be low if $\BE_5 - \BE_2 - \BE_3 + \TXbbd < m_\phi $ or $\BE_4 - 2 \BE_2 + \TXbbd < m_\phi$.} Since the shell model is fundamentally phenomenological and not predictive in terms of a UV completion, we cannot make definitive statements about the presence or absence of these bottlenecks.  We will thus frame our discussion of nugget synthesis in terms of the presence or absence of bottlenecks.

\section{Nugget Synthesis}
\label{sec:synthesis}

We have just reviewed the conditions for the formation of an $N=2$ nugget ($\DN[2]$).  Once a population of $\DN[2]$ exists, there are a number of routes to creating larger nuggets.  Here we discuss synthesis of larger-$N$ nuggets given either the presence or absence of low-$N$ bottlenecks. First we review fusion processes and reasonable models for their respective cross sections. Next we set up the Boltzmann equations that govern the evolution of the mass function and introduce a dimensionless interaction time---a dimensionless function of dark number density, typical cross section, typical speed, and Hubble rate---that sets the nugget size scale at the end of synthesis. Then we provide analytic estimates of typical nugget size after synthesis in both the presence and absence of bottlenecks. Finally we present semi-analytic solutions to the Boltzmann equations for the mass function after synthesis and relate them to our analytic estimates. Our semi-analytic discussion of the solution to the Boltzmann equations is similar to that of \cite{Hardy:2014mqa}, though with two significant differences.  
First, utilizing the results of our companion paper, we are able to compute the synthesized nugget mass in terms of the parameters of the model, $m_X$, $m_\phi$, and $\alpha_\phi$. Second, we employ the Compound Nucleus Model, and account for the possibility of reactions resulting in final states that include an extra nugget emission, which can lead to a broader mass function exiting synthesis.

\subsection{Dominant Processes and Cross Sections}\label{sec: dominant processes}

The relevant reactions, ignoring any contribution from $N$-body interactions for $N>2$, are of the form, 
\begin{align}
  \DN[i] + \DN[j] \longleftrightarrow \DN[k] + \DN[l] + \DN[m] + \ldots \label{eq: fusion}
\end{align}
where   
$\DN[0]$ denotes the mediator $\phi$, and $\DN$ number conservation dictates that $i+j = k+l+m+\ldots$. Here we describe and motivate using the Compound Nucleus (CN) model of nuclear physics for modeling cross sections for these processes. 

Within the CN model, one treats such reactions in two stages: first, an excited CN of size $i+j$ is formed, and then the compound state disintegrates \cite{1936Natur.137..344B} (see also {\em e.g.} \cite{weisskopf}). If any excess energy within the CN is quickly shared amongst all constituents, then it is reasonable to treat the two stages as independent processes. In this case, the probability for disintegrating into any given final state depends only on the energy, angular momentum, and parity of the CN, with no specific dependence on how the state is formed.  We can expect this so-called Bohr assumption to apply in most nugget reactions involving at least one saturated nugget in the early Universe. This is because the characteristic time for energy to be transferred across a nugget of size $R$ is order $R/v_s$ where $v_s = dp/d\epsilon$ is the speed of sound within a nugget, which we find to be a substantial fraction of the speed of light for saturated nuggets. On the other hand, the timescale over which the nuggets interact is order $R \sqrt{1-b/R} / v$, where $b$ is the impact parameter and $v$ the relative speed of the initial state nuggets. For large nuggets, $v$ is typically much smaller than the speed of light, thus we expect the excess energy to be efficiently distributed within the compound system over the time of a typical interaction as long as $b\lesssim R$. There is a potential exception coming from interactions with $b\simeq R$, where grazing interactions could lead to two-to-two dark number exchange processes. 

Given the Bohr assumption, the cross section for reactions \Eq{eq: fusion} takes the form
\beq
\sigma_{i j \rightarrow \alpha } = \sigma_{i j}  {\Gamma_{\alpha; i+j} \over\Gamma}  ,
\eeq  where $\sigma_{i j}$ is the cross section for forming the compound state of size $i+j$ and $\Gamma_{\alpha; i+j}$ is the rate for the compound state to disintegrate into a final state, $\alpha$. We will now address models for the formation cross section $\sigma_{i j}$ and the disintegration rates, $\Gamma_{\alpha; i+j}$, in turn.

For collisions of two large, saturated nuggets, we adopt a geometric cross section, 
\beq
\sigma_{i j} = \pi (R_i+R_j)^2 \qquad \text{where} \qquad R_i = i^{1/3} r_0
\label{eq:sat_radius}
\,
\eeq 
with $r_0$ given by \Eq{eq: r0 estimate}.
For collisions between a large saturated nugget and a much smaller nugget ($i \gg j$) we adopt a quantum-corrected geometric cross section similar to models used for neutron capture,

\beq
\sigma_{i j} = \pi (R_i+1/p)^2 \trans \qquad \text{where} \qquad \trans={4 p p' \over (p+p')^2} \qquad (i \gg j). \label{eq: capture cross section}
\eeq Here $p = \sqrt{E_j^2 -m^2}$ is the small nugget momentum and $p' = \sqrt{E_j^2 - m_\text{eff}^2}$ is the effective momentum inside the large nugget, and $m_\text{eff} = j \mu $ is the effective mass of the small nugget inside $\DN[i]$. The $1/p$ correction to the radius accounts for the effective de Broglie wavelength of the small nugget, which leads to an enhancement for small nugget capture. For larger nuggets sizes $R_i$, $1/p$ is typically sub-dominant and thus neglected. The transmission factor $\trans$ accounts for quantum reflection effects due to a sudden change in the small nugget's effective mass upon entering the large saturated nugget; note that in the limit that the smaller nugget is also saturated, $p = p'$ and thus $\trans = 1$.     

Time reversal invariance or reciprocity relates disintegration rates to formation cross sections for the reverse process. Consider the decay of a CN, $\DN[N]^*$, to a less excited state $\DN[N-k]^*$, while emitting a much smaller nugget $\DN[k]$ (or a mediator for $k=0$).  It can be shown (see Refs.~\cite{weisskopf,1937PhRv...52..295W}), when the density of states into which the CN can decay is large, that the partial width of the CN to decay into products $\DN[N-k]^* + \DN[k]$ is described by a thermal distribution weighted by the energy release (using the liquid drop model in Eq.~\ref{eq: liquid drop}) in the decay, 
\beq
\Gamma_k & = & 
\int \frac{g_k p^2}{2\pi^2}\sigma_k v\, e^{-\Delta E^*/T(E^*)} dp
\nonumber \\ 
& \approx & {g_k \over 2 \pi^2} e^{-(M_k-k \mu)/T(E^*)}  \int \sigma_{k} e^{-\epsilon/T(E^*)} \epsilon(\epsilon+ 2 M_k)d\epsilon\, ,
\eeq 
where $\Delta E^* \approx \sqrt{M_k^2 + p^2} - k \mu$ is the energy release. $T(E^*)$ is the temperature of the nucleons in the excited CN, $\epsilon \equiv \sqrt{M_k^2 + p^2} - M_k$ is kinetic energy of the emitted particle, $g_k$ the degrees of freedom of the emitted particle, and $\sigma_k$ is the cross section for the inverse process $\DN[N-k]^* + \DN[k] \rightarrow \DN[N]^*$. 
For very low temperatures satisfying $T \ll M_k - k \mu$, it is immediately apparent that emission of the lightest possible particles is heavily favored if $g_k \sigma_k$ is not radically different for different $k$. In our case, we will see that the temperatures are sufficiently small that we expect $\phi$ emissions to dominate. 

 
At low temperatures, the nugget is described as a Fermi gas with a modified fermion mass $m_*$. Then, to leading order in the excitation energy $E^*$, the temperature is given by ${T \sim \sqrt{E^*/N (4 \mu / \pi^2)} }$, where $\mu=m_{X 0}$ is the Fermi energy given by \Eq{eq: r0 estimate} . The binding energy release in the process, $\DN[f N] + \DN[(1-f) N] \rightarrow \DN[N]$, is given by $Q = \epsilon_\text{surf} N^{2/3} (f^{2/3}+(1-f)^{2/3} - 1)$ , which is maximal when $f=1/2$ for fixed $N$. Thus, assuming negligible kinetic energy in comparison to binding energy release, the maximal excitation energy of a CN is $E^* \sim 0.26\, \epsilon_\text{surf} N^{2/3}$. This leads to maximum temperatures of order 
$
T_{\max} \sim 0.3 N^{-1/6} \sqrt{\epsilon_\text{surf} \mu} \sim 0.4 (N / N_\text{sat})^{-1/6} \sqrt{\epsilon_\text{surf} m_\phi},
$
where in the final expression we used the definition $N_\text{sat} \equiv (r_0 m_\phi)^{-3}$ and \Eq{eq: r0 estimate}. In Ref.~\cite{Gresham:2017zqi}, we found that $\epsilon_\text{surf}$ is order $3\, m_X$ to $10\, m_X$ for sample parameters $\alpha_\phi =0.1, 0.01$ and $m_\phi/m_X = 10^{-2}, 10^{-3}, 10^{-4}$, and it mildly increases with decreasing $m_\phi/m_X$ and decreasing $\alpha_\phi$.  At saturation, the typical temperature of the formed CN will be much smaller than $m_X$ but possibly comparable to or larger than $m_\phi$. 
In the parameter ranges of interest (where saturation applies), scalar emissions generally dominate by many orders of magnitude over small nugget emission, with stronger domination as $N$ grows and $m_\phi/m_X$ falls. 

Despite domination of scalar emission over small nugget emission at each step in the cascade, one might worry that the large number of steps in a CN cascade could lead to a sizable total small nugget emission rate. An upper bound on the number of $X$ emissions can be estimated as $N_X  \lesssim N_\phi \Gamma_X(E^*_0)/\Gamma_\phi(E^*_0)$ with $N_\phi$ the number of $\phi$ emissions given an initial excitation energy $E^*_0$. Using the further upper-bound estimate $N_\phi \lesssim E^*_0/m_\phi$, we find that $N_X$ is substantially below one for the relevant parameter ranges discussed in Sec.\ref{sec: synthesis start conditions}.

\subsection{Boltzmann Equations}
\label{sec:boltzmann_equations}
We have just seen in Sec.~\ref{sec: dominant processes} that \emph{coagulation}---fusion through emission of mediators only---will typically dominate when the Bohr assumption of quick thermalization applies.  At larger impact parameter, a thermal model may not apply and nugget fragmentation could occur. We will suppose that---if they are relevant at all---the dominant form of non-thermal fusion reactions are of the form $\DN[i] + \DN[j] \rightarrow \DN[i+j-k]+\DN[k>0]+\phi$s, which we will refer to as \emph{two-to-two} reactions.  

In either case, the nugget distribution will follow a set of Boltzmann equations. Defining $y_k \equiv n_{\DN[k]}/n_X$ with $n_k$ the number density of a nugget of size $k$ and $n_X = \sum_k k n_k$, we have
\begin{multline}
{d y_k  \over d t} = n_X \bigg( \sum_{i \le j} y_i y_j \langle \sigma v (\DN[i]+\DN[j] \rightarrow \DN[k]+\DN[i+j-k])\rangle \\
- y_k \sum_{m< n} y_{m+n-k} \langle \sigma v (\DN[k]+\DN[m+n-k] \rightarrow \DN[m]+\DN[n]) \rangle \bigg)\,,
\label{eq:boltzmann_t}
\end{multline}
and for notational simplicity, we denote coagulation as a two-to-two process with a final state particle $\DN[0]$.  All fusion processes will generally include many additional $\phi$ emissions, which do not impact the nugget number distribution. At low temperature, only exothermic processes contribute, and the summations in \Eq{eq:boltzmann_t} are restricted to final states that are more asymmetric than the initial state. 

Total dark number conservation implies that $ \sum_k k  y_k=1$, so that in the large $N$ continuum, $k y_k$ becomes a probability distribution. There are two sources of time dependence for the Boltzmann equation: the density $n_X $, which dilutes away as $\sim 1/a^3$, and the thermal-averaged cross sections, which may include a nontrivial transmission factor $\trans$. It is possible to factor out the time dependence by defining a dimensionless interaction time $\gamma$, such that (c.f.~$w$ in \cite{Hardy:2015boa}),
\beq
{d \gamma \over d t} = n_X(t) \sigma_\circ \langle v \trans \rangle_\circ (t)\, , \label{eq: gamma}
\eeq 
where $\sigma_\circ = \pi r_0^2$, and $\langle v \trans \rangle_\circ$ is the common factor obtained from a thermal average of the velocity-dependent part of the cross sections. For fusion of nuggets both deep in the saturation regime, $\trans=1$, and we simply take $\langle v \trans \rangle_\circ \simeq \sqrt{T_X/\bar m_X}$. Note that the binding energy per particle can be very large and thus $\bar m_X$ can be significantly smaller than $m_X$. For $\DN[2]$-large fusion, relevant in bottleneck scenarios, we instead take $\trans \sim 4 v$, and $\langle v \trans \rangle_\circ \simeq 4{T_X/ m_X}$. In terms of $\gamma$, the Boltzmann equation becomes
\begin{multline}
{d y_k  \over d \gamma} = \bigg( \sum_{i\le j} y_i y_j {\langle \sigma v (\DN[i]+\DN[j] \rightarrow \DN[k]+\DN[i+j-k])\rangle \over \sigma_\circ \langle v \trans \rangle_\circ } \\
- y_k \sum_{m< n} y_{m+n-k}  { \langle \sigma v (\DN[k]+\DN[m+n-k] \rightarrow \DN[m]+\DN[n]) \rangle \over \sigma_\circ\langle v \trans \rangle_\circ } \bigg)\, ,
\label{eq:boltzmann}
\end{multline}
where the factor $\sigma_\circ\langle v \trans \rangle_\circ$ absorbs the time dependence of $\langle \sigma v \rangle$, such that the temporal evolution of the nugget distribution is entirely captured by $\gamma(t)$. 

In general, the nugget bath temperature $T_X$ can deviate from the temperature of the photon bath $T_\gamma$, if the dark sector is not in kinetic equilibrium with the SM. Depending on whether the nugget bath is relativistic ($T_X \propto a^{-1}$) or not ($T_X \propto a^{-2}$), the form of $\gamma(t)$ varies. In the deep saturation limit during radiation domination, one has
\beq
\gamma = {n_X(t_{\rm syn}) \sigma_\circ  \over H(t_{\rm syn})}\sqrt{T_{X}^{\rm syn} \over \bar m_X} \left\{ 
\begin{array}{l l} 
 {1 \over 2}  \left[ 1- \left( a \over a_{\rm syn} \right)^{-2} \right], \qquad (T_X \propto a^{-2}) 
\vspace{1mm}
\\
{2 \over 3}  \left[ 1- \left( a \over a_{\rm syn} \right)^{-3/2} \right],  \qquad (T_X \propto a^{-1})
\end{array}\right. ,
\label{gammaparam}
\eeq
and for case of $\DN[2]$-large fusion, one has
\beq
\gamma = {n_X(t_{\rm syn}) \sigma_\circ  \over H(t_{\rm syn})}{4 T_{X}^{\rm syn} \over m_X} \left\{ 
\begin{array}{l l} 
{1 \over 3}\left[    1- \left( a \over a_{\rm syn} \right)^{-3} \right], \qquad (T_X \propto a^{-2}) 
\vspace{1mm}
\\
{1 \over 2}  \left[ 1- \left( a \over a_{\rm syn} \right)^{-2} \right],  \qquad (T_X \propto a^{-1})
\end{array}\right. ,
\label{gammaparam_bottleneck}
\eeq
where the subscript ``syn'' denotes the value of the quantity at the beginning of synthesis. For $\BE_2$ not radically greater than $m_\phi$, $\phi$ is nonrelativistic at $\Tbbd$ already, and without an additional dark relativistic component, the dark bath will be nonrelativistic. 

As discussed later, the typical nugget size exiting synthesis will be of order $\gamma_\text{max}^{6/5}$ with $\gamma$ determined through \Eq{gammaparam} in the absence of a bottleneck, and the typical size of a subdominant population of very large nuggets will be order $\gamma_{\text{max}}^{3}$ with $\gamma$ determined through \Eq{gammaparam_bottleneck} when a strong bottleneck is present.

\subsection{Large Nugget Formation in the Absence of a Bottleneck}
\label{eq:nugget_formation_noBN}

%

Here we derive an analytic understanding of the typical size of a nugget exiting synthesis. We start by defining an average nugget size. Motivated by the fact that $\sum_k k y_k=1$, which indicates $k y_k$ acts like a probability distribution in the continuum limit, we define the average size by $\bar k \equiv \sum_{k=1} k^2 y_k$.\footnote{An alternative definition $\bar k = {\sum_{k=1} k y_k  \over \sum_{k=1} y_k} = (\sum_k y_k)^{-1}$ corresponding to a number density-weighted average is not appropriate when two-to-two processes contribute significantly toward fusion of large nuggets. This is because $\sum_k y_k$ does not change under two-to-two processes, but only under fusion through mediator emissions.} For large $k$, $m_k \propto k \bar m_X$, and $\bar k$ is essentially an energy density-weighted average. The evolution of $\bar k$ follows 

\begin{align}
 \frac{d \log\bar k}{d\gamma} = \bar k^{-1} \sum_{k<i\le j} 2y_i y_j (i-k)(j-k) 
{\langle \sigma v (\DN[i]+\DN[j] \rightarrow \DN[k]+\DN[i+j-k])\rangle \over \sigma_\circ \langle v \trans \rangle_\circ }\,.
\label{eq:boltz_kbar}
\end{align}
Here the summation only includes processes where the final states have a more asymmetric nugget distribution, such that ${i+j-k>j\ge i >k}$, implying that $\bar k$ is monotonically increasing. 

To proceed further, we take the saturation limit, which will be a good approximation as long as $\bar k \gtrsim N_\text{sat}$. Consider the most dominant contributions in Eq.~\ref{eq:boltz_kbar}, which comes from fusions of large nuggets with sizes around $\bar k$, and emissions (mostly) of mediator particles $\phi$. This process roughly doubles the size, and leads to an ${\cal O}(1)$ change in $\log \bar k$. When the time scale for this process to occur exceeds $\gamma_{\rm max}$, synthesis freezes out. Therefore, we approximate Eq.~\ref{eq:boltz_kbar} by replacing the cross section (summed over $k$) in Eq.~\ref{eq:boltz_kbar} by the total $\bar k{\rm -}\bar k$ interaction rate, and substitute 
$(k-j)(k-i)\sim \bar k^2$
and $\sum_{i,j}y_i y_j \sim 1/\bar k^2$. The freeze-out condition then becomes 

\beq
{d \log \bar k  \over d \gamma} \sim 
{\bar k}^{-1}  { \langle \sigma v (\DN[\bar k]+\DN[\bar k] \rightarrow \DN[\sim 2 \bar k]) \rangle \over \sigma_\circ \langle v \trans \rangle_\circ } \lesssim \frac{1}{\gamma_{\rm max}}. 
\eeq

In the saturation limit, the cross section for $\bar k {\rm -} \bar k$ fusion scales like $\sigma_{\bar k \bar k}\sim \sigma_\circ \bar k^{2/3} $, with a velocity dependence $v_{\bar k \bar k}\sim v_\circ \bar k^{-1/2}$. (The transmission factor is $\trans =1$ in this limit.) Then, setting  $\langle \sigma v (\DN[\bar k]+\DN[\bar k] \rightarrow \DN[\sim 2 \bar k]) \rangle = \sigma_\circ v_\circ \bar k^{1/6}$, 
the typical nugget size at the end of synthesis is 
\beq
\bar k_{\rm fo} \sim  \left( {\gamma_\text{max}} \right)^{6/5}, \label{eq: kbar estimate}
\eeq
with $\gamma_{\max}$ given by \Eq{gammaparam} when $a_\text{syn}/a \rightarrow 0 $, and the typical nugget mass is $\bar M_\text{fo} \sim \bar k_{\rm fo} \bar m_X$.
Explicitly,
\begin{multline}
\gamma_\text{max} \approx {10^{6}} \sqrt{g_{* S}^\text{syn} \over g_*^\text{syn}} \sqrt{g_{* S}^\text{syn} \over 10 } {\Tgambbd \over  \TXbbd} \left(\TXbbd \over \BE_2 / 28 \right)^{3/2} \\
\times  \left( r_0 m_X \over 23 \right)^2 \left(10 \text{GeV} \over m_X \right)^2 \left(400 \BE_2 \over  m_X \right)^{3\over 2} \left( m_X \over 10 \bar  m _X \right)^{3/2},
 \label{eq: gamma 2}
\end{multline}
where fiducial values correspond roughly to the parameters $\alpha_\phi = 0.1$, ${m_X = 10\, \text{GeV}}$ and ${m_\phi = 10\, \text{MeV}}$. 

\begin{figure}
\begin{minipage}{0.5\textwidth}
\includegraphics[width=1\textwidth]{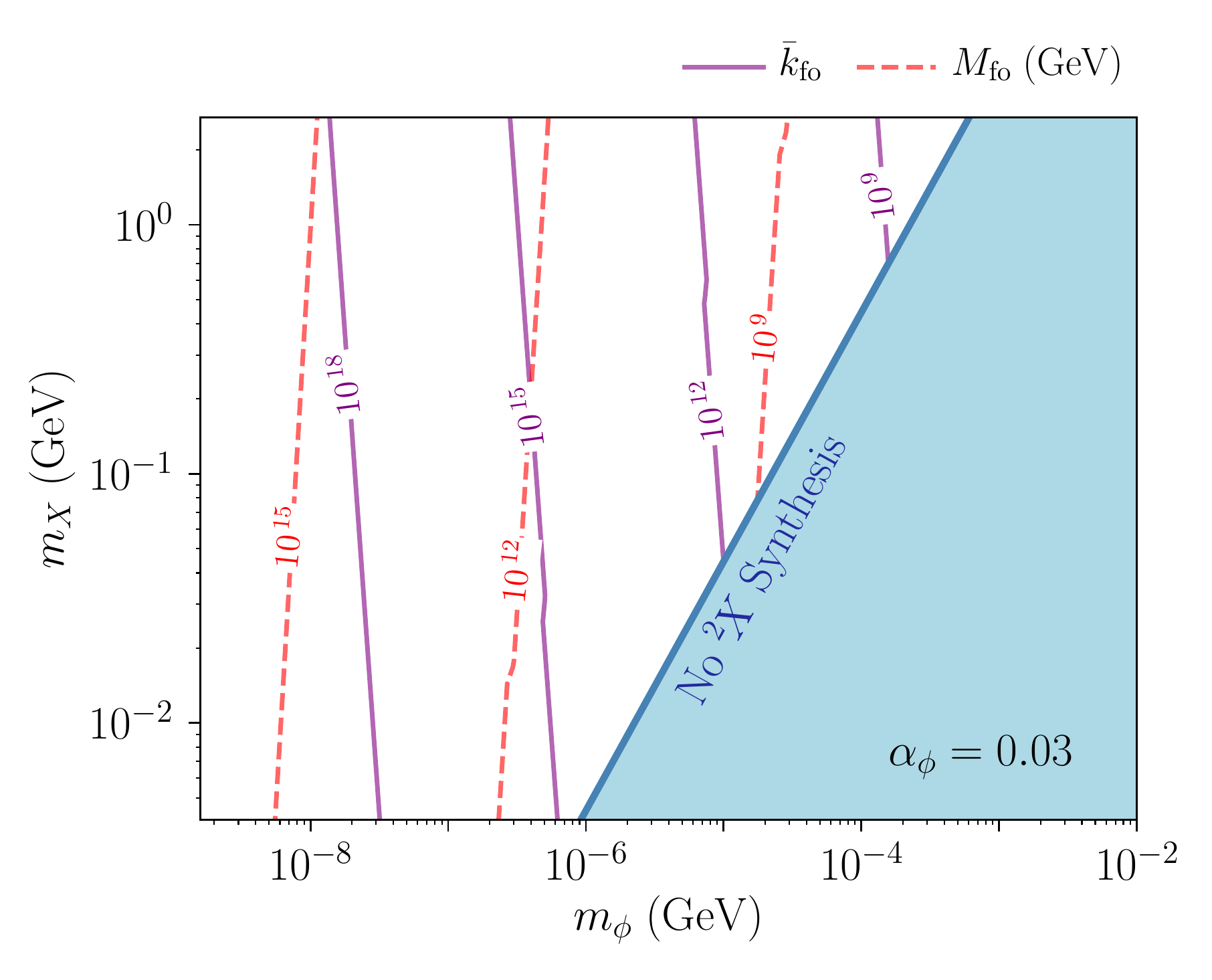}
\end{minipage}
\begin{minipage}{0.48\textwidth}
\includegraphics[width=1\textwidth]{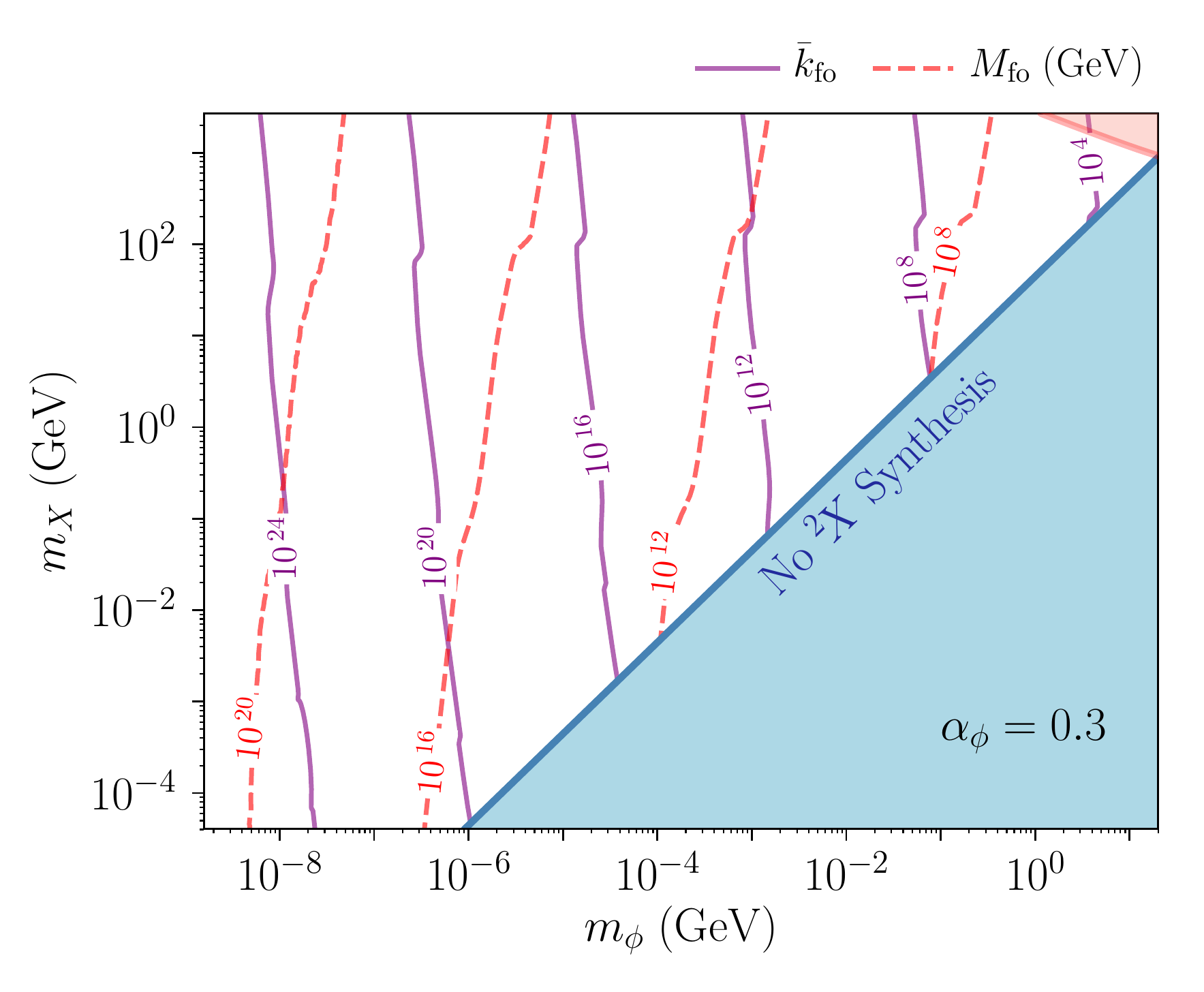}
\end{minipage}
\caption{
Contours of typical nugget number exiting big bang darkleosynthesis, $\bar k_\text{fo}$ (\emph{dashed red}) and typical nugget mass $\bar M_{\text{fo}}$ (\emph{solid purple}) for $\alpha_\phi=0.03$ (left) and $\alpha_\phi=0.3$ (right). The temperature of the dark sector is assumed to be roughly $T_X \approx T_\gamma$. The blue shaded region corresponds to $\BE_2<m_\phi$, where $\DN[2]$ synthesis will not be efficient (\Eq{eq: synthesis condition 1}). The upper $m_X$ cutoff corresponds to the requirement that $\DN[2]$ fusion rate is smaller than Hubble as in \Eq{eq: synthesis condition 2}, and the lower $m_X$ cutoff corresponds to $T_\text{synth} \lesssim 10 T_\text{eq}$.  (See Fig.~\ref{fig: synthesis conditions}.). The various kinks in the contours are results of the change in $g_*$ as the synthesis temperature passes through QCD phase transition and neutrino decoupling.}
\label{fig: Nmax}
\end{figure}

In Fig.~\ref{fig: Nmax} we show contours of constant $\bar k_\text{fo}$ and $\bar M_\text{fo}/\text{GeV}$ assuming saturation values for $r_0$ and $\bar m_X$ as described in \cite{Gresham:2017zqi} and summarized in Sec.~\ref{sec: saturation properties}, as a function of $m_X$ and $m_\phi$, with two choices of fixed $\alpha_\phi$. We have taken $\Tgambbd = \TXbbd$, with $\TXbbd$ estimated as described in \S \ref{sec: synthesis temperature}, and we have included the contribution of $\phi$ to $g_*$ when relevant.\footnote{As discussed in Sec.~\ref{sec: synthesis temperature}, depending on details of the cosmology, we expect $\Tgambbd = \TXbbd$ to within a factor of a few, depending on the decoupling time of the two sectors.} The $m_X$ range shown is that which satisfies the synthesis condition \Eq{eq: synthesis condition 2} and the condition $\TXbbd > 10 T_\text{eq}$. The shaded region at smaller $m_X/m_\phi$ does not satisfy the synthesis condition in \Eq{eq: synthesis condition 1}.   We chose the lower cutoff for $m_\phi$ according to $\sqrt{m_\phi m_X} \gtrsim 10^{-8}$GeV. For smaller values of $\sqrt{m_\phi m_X}$, as discussed in Sec.~\ref{sec: synthesis temperature}, if the SM and DM sectors are kinetically decoupled and the thermal DM sector does not contain additional light degrees of freedom, the dark sector temperature could increase by an order of magnitude or more during synthesis, which would increase the estimate by a similar factor. 

In Fig.~\ref{fig: Nmax}, we can see that $\bar k_\text{fo}$ and $\bar M_\text{fo}$ depend much more strongly on $m_\phi$ and $\alpha_\phi$ than on $m_X$. Using $\BE_2 = \alpha_\phi^2 m_X /4$ and the estimates \Eq{eq: r0 estimate}, modulo the very weak dependence of $g_*$ and $\BE_2/\TXbbd$ on model parameters, we find that the typical nugget number $\bar k_\text{fo}$ and mass $\bar M_\text{fo}$ scale as 
\beq 
\bar k_\text{fo} \propto \alpha_\phi^{93/20} m_X^{-3/10} m_\phi^{-21/10}  \qquad \text{and} \qquad \bar M_\text{fo} \propto \alpha_\phi^{22/5} m_X^{1/5} m_\phi^{-8/5}.  \label{eq: no bottleneck scaling}
\eeq 
This scaling is reflected in the plot. For example, it is worth noting that $\bar k_\text{fo}$ grows slightly more rapidly with decreasing $m_\phi$ than does $\bar M_\text{fo}$. This is entirely due to the increase of binding energy per particle with decreasing $m_\phi$.

Recall that our estimate is valid only when $\bar k_\text{fo} > N_\text{sat}$ since we have used the saturation cross section and binding energy in our freeze-out estimate. In the strong binding limit, 
\beq
N_\text{sat} \equiv (m_\phi r_0)^{-3} \approx \alpha_\phi^{-3/4} \left({m_X \over m_\phi}\right)^{3/2}\,.
\eeq
Thus since $\bar k_\text{fo}$ scales more strongly with $m_\phi^{-1}$, for fixed $\alpha_\phi$ and $m_X$, the approximation becomes better as $m_\phi$ decreases. 
Numerically, we find that for $\alpha_\phi \lesssim 0.1$, the self-consistency condition for applying the saturation limit $\bar k_\text{fo} > N_\text{sat}$ is always satisfied whenever the synthesis conditions are met. And in Fig.~\ref{fig: Nmax}, where the region with $\bar k_\text{fo} < N_\text{sat}$ for $\alpha_\phi = 0.3$ is shaded red, we see that even for larger $\alpha_\phi$ the estimate is invalid only in a small region of parameter space.

\subsection{Nugget Distribution Function in the Absence of a Bottleneck}
\label{sec:nuggetmassfunction}

So far we have focused on order of magnitude analytic estimates. Here we support and complement our estimates with a full semi-analytic analysis of the Boltzmann equations in Eq.~\ref{eq:boltzmann}, complemented with numerical calculations. As before, we only consider coagulation and two-to-two processes. Although much of the analysis carries over when higher order processes are included, as long as they stay homogeneous with the same weight as we will discuss.
It is useful to rewrite the infinite set of differential equations in Eq.~\ref{eq:boltzmann} by introducing a kernel, $K(i,j,k)$,
\begin{align}
  K(i,j,k)={\langle \sigma v (\DN[i]+\DN[j] \rightarrow \DN[k]+\DN[i+j-k])\rangle \over \sigma_\circ\langle v \trans \rangle_\circ }
\end{align}
so that the Boltzmann equation, \Eq{eq:boltzmann_t}, becomes
\begin{align} 
\frac{dy_k}{d\gamma}=\left[
 \sum_{i\le j} y_i y_j K(i,j,k)
- \sum_{k+l< 2m}y_k y_{l} K(k,l,m)
\right]\,.
\label{eq:boltz_kernel}
\end{align}
Analogous equations have been considered in the statisitical and mathematical physics literature (see Ref.~\cite{2010kvsp.book.....K} for a pedagogical introduction), and when $K(i,j,k)\propto \delta_{i+j,k}$, \Eq{eq:boltz_kernel} is known as the Smoluchowski equation for coagulation \cite{1916ZPhy...17..557S}. Here we consider the saturation limit and utilize the CN-like picture for two-to-two processes, such that the kernel scales simply as,
\begin{align}
  K(i,j,k) =\sqrt{\frac{1}{i}+\frac{1}{j}} \left(i^{\frac{1}{3}}+j^{\frac{1}{3}}\right)^2 \frac{\Gamma_{k}}{\Gamma}\, , \label{eq: kernel}
\end{align} 
where $\Gamma_{k}$ is proportional to the partial width of a compound state $i+j$ transitioning into a final state $k + (i+j-k)$, the squared factor characterizes the scaling of the geometric cross section, and the square root factor characterizes the relative speed. 
A similar kernel was considered in \cite{Hardy:2014mqa}, but with $\Gamma_{k}=\delta_{i+j,k}$, corresponding to the case of coagulation \cite{1916ZPhy...17..557S}. There are generally no closed form solutions even for a simplified choice of fusion kernel \cite{2010kvsp.book.....K}, and given that $\bar k$ and the total interaction time $\gamma_{\rm max}$ are typically very large, numerical calculations quickly become intractable. Fortunately, as similarly discussed in \cite{Hardy:2014mqa}, 
when the kernel is homogeneous $K(i,j,k)= \lambda^{-\alpha}K(\lambda  i,\lambda j,\lambda k)$, scaling solutions exist, which allow for extrapolation of numerical results for small $\gamma$ to large $\gamma \sim \gamma_{\rm max}$. In the saturation limit, the total inelastic cross-section is already homogeneous with degree $1/6$. Then a scaling solution is valid as long as $\Gamma_k$ in \Eq{eq: kernel} is also homogeneous, which we will assume for now. It is important to note that the kernel, \Eq{eq: kernel}, applies only when large saturated nuggets dominate the dark matter density; for fusion involving at least one unsaturated nugget, the form of the kernel will change, and for the most general fusion interactions, one cannot even define a time-independent kernel. We will check that our results are self-consistent in that the vast majority of nuggets exiting synthesis are above the saturation size, where the kernel we use does apply.

We will now derive the scaling solution, following \cite{2010kvsp.book.....K} in spirit. In the large $\bar k$ limit,  nugget indices may be treated as continuous variables. Consider the ansatz $y_k(\gamma) =s^2 f(k s) $, where $s(\gamma)$ is some function of $\gamma$ that is to be determined. Substituting the ansatz into Eq.~\ref{eq:boltz_kernel} and replacing summation with integration in the continuum limit, we have 
\begin{align}
\dot s s \bigg[ (ks) f'(k s) + 2 f(k s) \bigg]
= 
 s^{2-\alpha}\bigg[ &\iint d(i s) d(j s) K(i s, j s, k s) f(i s)f(j s) \notag \\
&
-\iint d(l s) d(m s) \, K(k s ,l s, m s) f(k s)f(l s)
\bigg]\,,
\end{align}
where we have used the homogeneity property of $K$ to change the integration variable. We have not explicitly included the integration bounds, which do not affect the derivation as long as they are linear functions of the integration variables.  One sees that for $s = c \gamma^{1/\alpha}$, the $\gamma$ dependence drops out entirely, and one is left with an integro-differential equation for $f(x)$, given by
\begin{align}
xf'(x) + 2 f(x)
=  \alpha c^{-\alpha} \bigg[\iint  dy dz\, K(y ,z, x) f(y)f(z)
 -\iint dy  dz \, K(x ,y,z) f(x)f(y)
\bigg]\,.
\label{eq:ODE_f}
\end{align}
For nugget fusion, we will consider $K$ with homogeneous weight $\alpha = -5/6$, and hence $s \sim \gamma^{-6/5}$. Eq.~\ref{eq:ODE_f} in general is still very difficult to solve even numerically. However, it is known that the nugget distribution generally approaches the scaling solution very quickly independent of the initial condition (see \cite{2010kvsp.book.....K}). Therefore, it is possible to numerically integrate Eq.~\ref{eq:boltz_kernel} by truncating the differential equation at finite nugget number, and then extract the scaling function $f(x)$ by testing that the solutions $y_k(\gamma)$ have converged to a scaling limit. 

It is illuminating to revisit our earlier discussion of typical nugget size $\bar k$ in Sec.\ref{eq:nugget_formation_noBN} in light of the scaling solution. In the scaling limit, $\bar k = s^{-1} \int dx \, x^2 f(x)$. Given that there is freedom to choose $f(x)$ by rescaling $s$, we may set  $\bar k = s^{-1}$. Then the scaling limit simply becomes
\begin{align}
  k^2y_k(\gamma) \rightarrow \frac{k^2}{\bar k^2}f\left( \frac{k}{\bar k} \right)\,.
\label{eq:scaling_solution}
\end{align}
In the scaling limit, $k^2 y_k$ maintains the same shape, centered on $\bar k$, with a scaling behavior ${\bar k \sim \gamma^{6/5}}$, verifying our earlier estimate. For our numerical study, we consider three separate homogeneous $\Gamma_k$ for the kernel $K(i,j,k)$. The three branching ratio forms we consider correspond to 
\begin{enumerate}
\item Coagulation: where $\Gamma_{i j \rightarrow k} \sim \delta_{i+j,k}$ 
\item Uniform: $\Gamma_{i j \rightarrow k l} \sim \theta(|k-l|-|i-j|)$. 
\item Energy Scaling: $\Gamma_{i j \rightarrow k l}\sim Q_{ij\rightarrow k l}^2 \sigma_{k+l}$, where the heat release $Q_{i j \rightarrow k l}$ is proportional to the change in total surface area for a given reaction. Specifically, $\Gamma(i+j\rightarrow k+l)\sim [(i^{\frac{2}{3}}+j^{\frac{2}{3}})-(k^{\frac{2}{3}}+l^{\frac{2}{3}})]^2 (k^{\frac{1}{3}}+l^{\frac{1}{3}})^2$, which roughly captures the increase in phase space as the reaction becomes more exothermic. 
\end{enumerate}

\begin{figure}[th]
\begin{minipage}{.45\textwidth}
\includegraphics[trim={0cm 0cm 0cm 0cm},clip,width=1.0\linewidth]{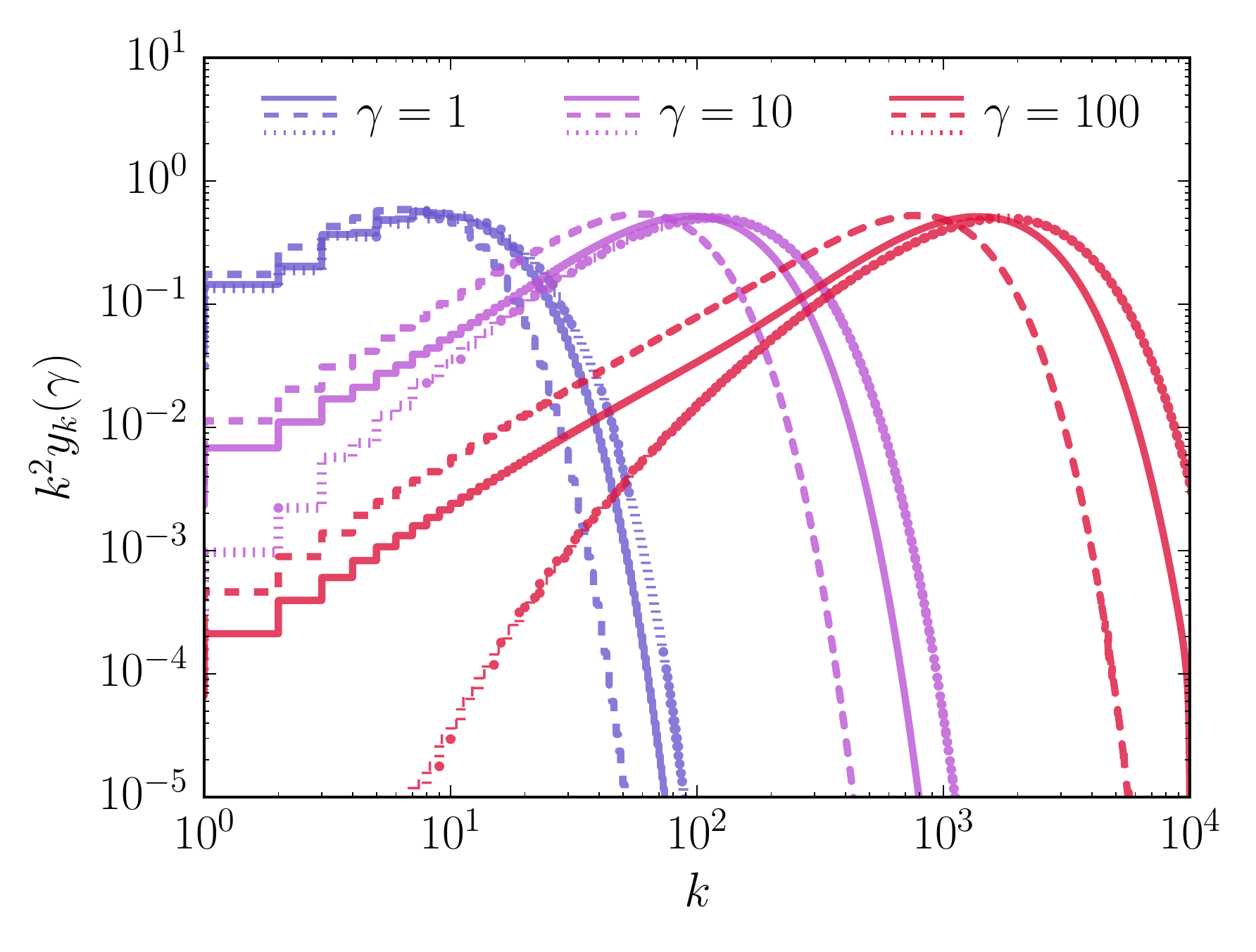}
\end{minipage}
\hspace{.3cm}
\begin{minipage}{.45\textwidth}
\includegraphics[trim={0cm 0cm 0cm 0cm},clip,width=1.0\linewidth]{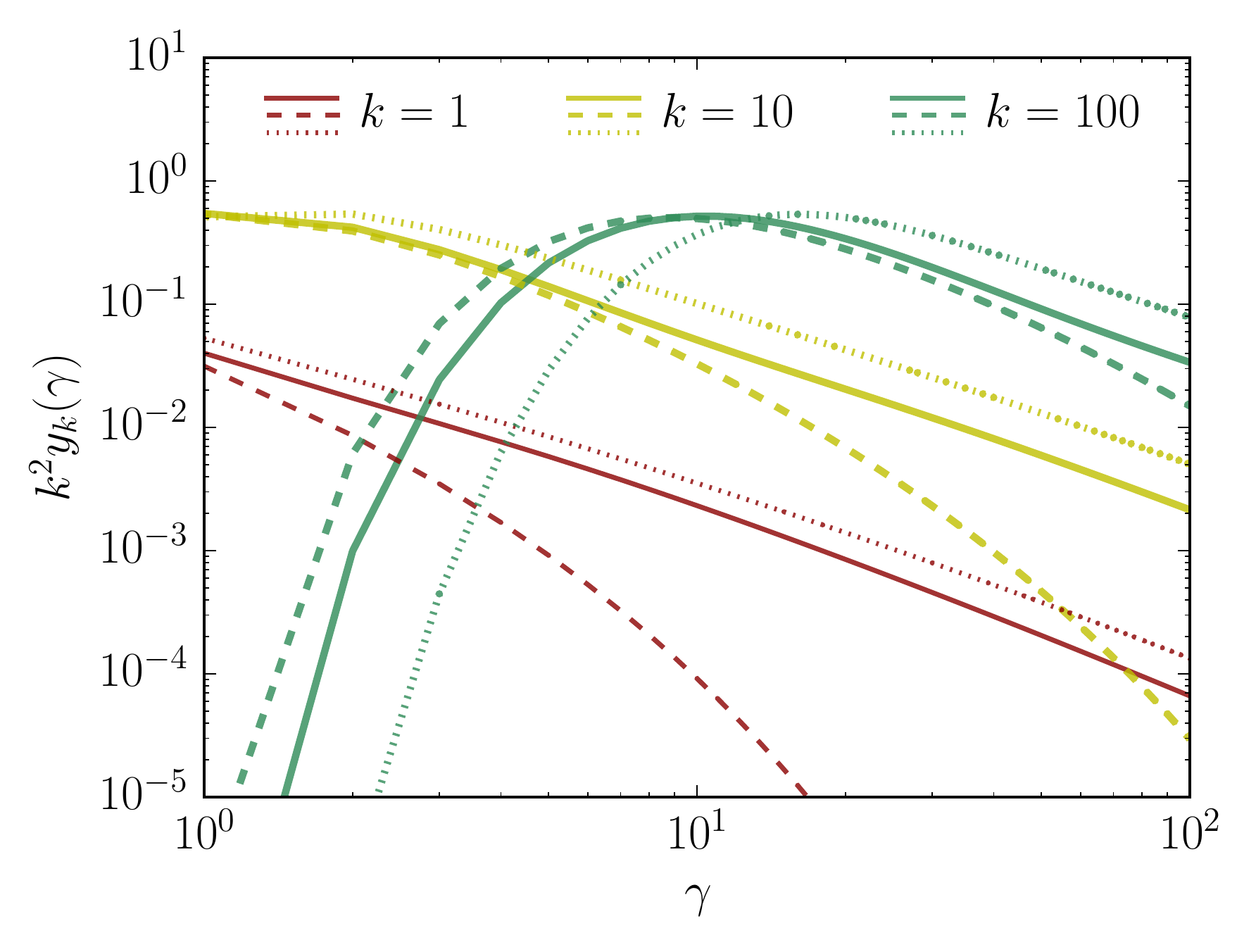}
\end{minipage}
\caption{Nugget distribution function at different interaction time $\gamma$ (right) and specific number fraction as a function of $\gamma$. Here, the different line style depicts different assumptions for the branching ratio, with solid/dashed/dotted corresponding to the Energy-Scaling/Uniform/Coagulation branching ratio assumption as described in the main text.}
\label{fig:synth_fusion}
\end{figure}

\begin{figure}[th]
\begin{minipage}{0.47\textwidth}
\includegraphics[width=1\textwidth]{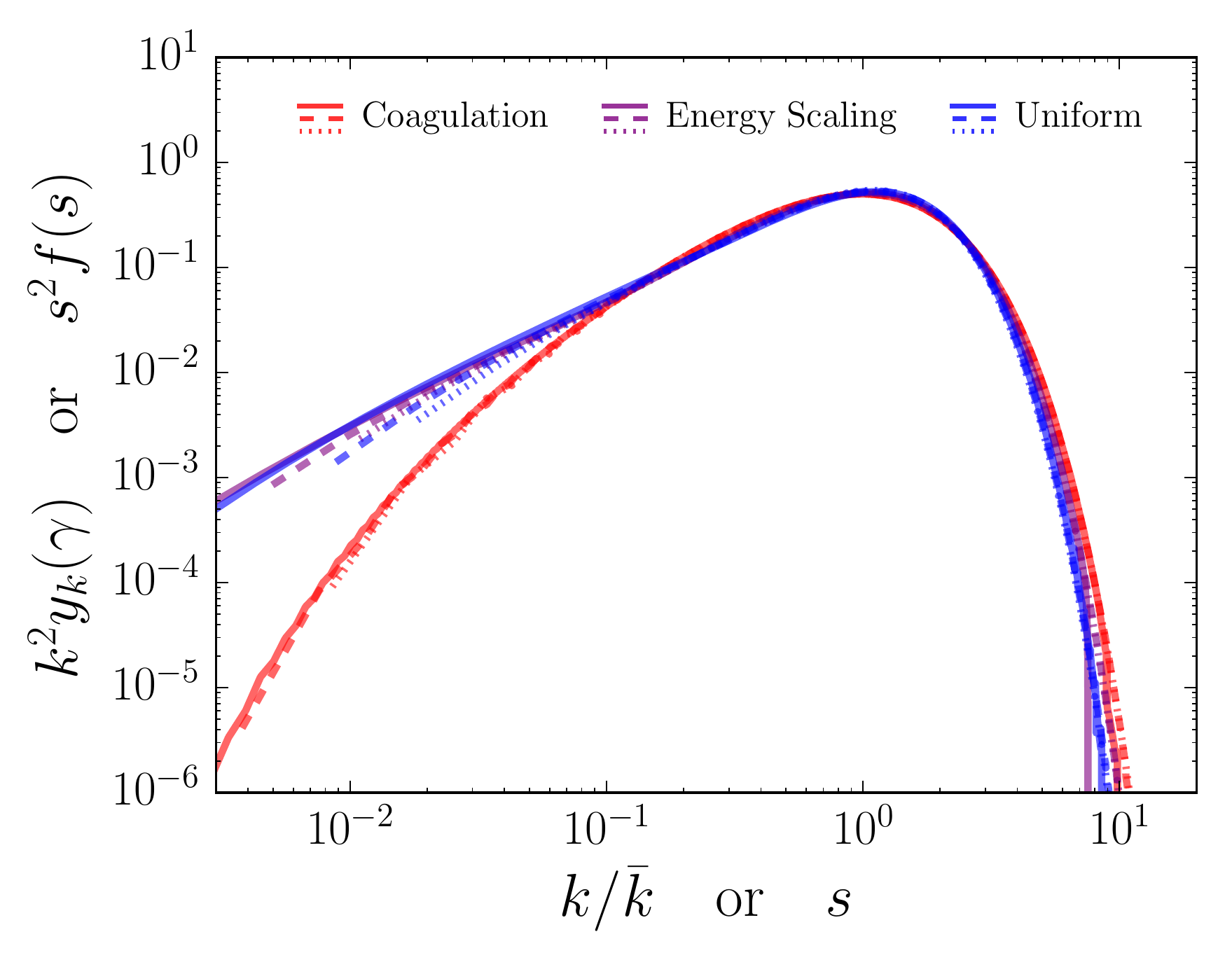}~
\end{minipage}
\begin{minipage}{0.51\textwidth}
\includegraphics[width=1\textwidth]{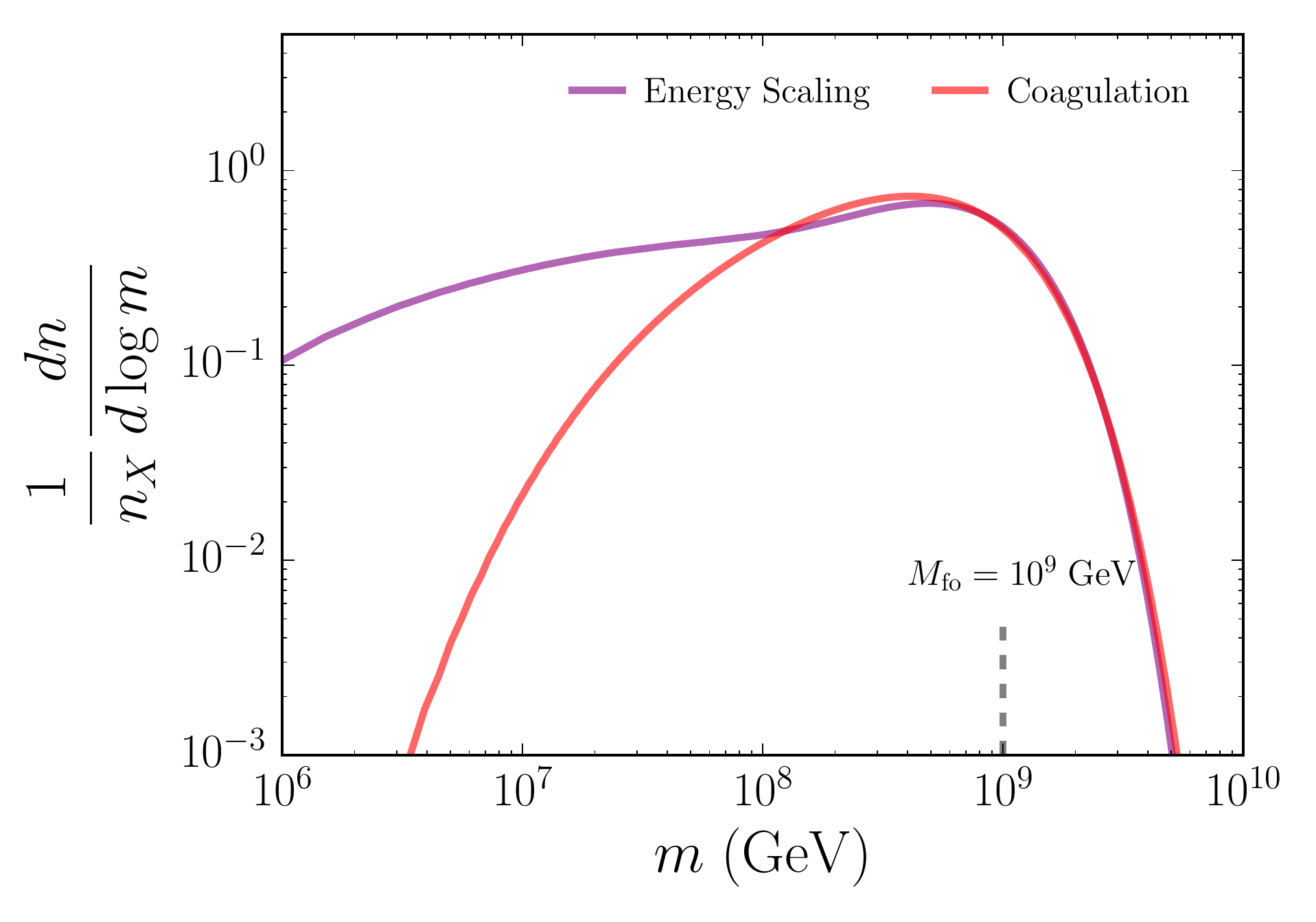}
\end{minipage}
\caption{(Left) Nugget mass functions $k^2y_k(\gamma)$ for different $\gamma$ and kernel assumptions. The different colors, red/purple/blue show the mass functions for the fusion/energy scaling/uniform assumptions for the kernel respectively. The solid/dashed/dotted line shows the curves for $\gamma=10,20,100$ respectively. All lines with the same color are seen to converge to a common function $s^2f(s)$ accordingly to \Eq{eq:scaling_solution}. The blue (uniform) and purple (energy scaling) curves converge to the same function approximately. (Right) The extracted differential nugget density exiting synthesis for $\bar M_{\rm fo}=10^9 $ GeV (the distribution is proportional to $s f(s)$). A different $\bar M_{\rm fo}$ simply shifts the $x$-axis logarithmically while maintaining the shape of the differential distribution. }
\label{fig:mass_function}
\end{figure}

 Starting with the initial condition $y_i(0)=\delta_{i0}$, Figure~\ref{fig:synth_fusion} shows $k^2 y_k(\gamma)$ as a function of nugget number $k$ and a specific nugget branching fraction as a function of interaction time $\gamma$. In the continuum limit, $k^2 y_k(\gamma)$ is interpreted as the differential nugget probability distribution, $dp(k)/d\log k$, or the fraction of the nugget number shared by nuggets centered on $k$ within a unit bin in $\log k$. We see that the nugget distributions very quickly become dominated by large $k$ even at small interaction time $\gamma=100$. Differences in the branching ratios do not significantly alter the behavior, although a flatter branching ratio tends to  enhance the nugget distribution at small $k$.

Fig.~\ref{fig:mass_function} shows the extracted mass function at different interaction time $\gamma \in \{10, 20, 100\}$, in dotted/dashed/solid line for the different branching ratio assumptions. One already sees a convergence to a universal function. At smaller nugget number, the mass functions are significantly broader for branching ratios that include two-to-two processes. Although the difference between the uniform and energy scaling branching ratios are small.

\subsection{Large Nugget Formation in the Presence of Bottleneck -- Nugget Capture}
\label{sec:bottleneck}

In this section we consider nugget synthesis in the presence of a bottleneck. As discussed in Sec.~\ref{sec:bottleneck}, we expect the nugget distributions to be impacted significantly only when both $\DN[3]$ and $\DN[4]$ are unstable. In this case, the two-to-two fusion processes are halted until larger nuggets ($\DN[5]$ or $\DN[6]$) are formed through other processes. For instance, $\DN[6]$ can be produced via rare three-$\DN[2]$ fusions, where a short-lived state $\DN[4]$ may exist, permitting the reaction ${\DN[4]+\DN[2]\rightarrow \DN[6]+\DN[0]}$ to proceed before the $\DN[4]$ decays. This is analogous to $3 \,{}^4\text{He}\rightarrow  {}^{12}\text{C}$  through the ${}^8\text{Be}$ resonance. In this case, the $\DN[6]$ now functions as a nucleation site, capturing nearby $\DN[2]$.


In the limit that the nucleation sites are sparse, the process $\DN[N]+\DN[2]\rightarrow \DN[N+2] $ with $\phi$ emissions dominates. Considering only this coagulation process, for large $k$, the Boltzmann equation in \Eq{eq:boltzmann} becomes
\begin{align}
  \frac{dy_k}{d\gamma}= y_2 \left(
y_{k-2} \frac{\langle \sigma v (\DN[k-2]+\DN[2] \rightarrow \DN[k])\rangle}{\sigma_\circ\langle v \trans \rangle_\circ}
  - y_{k} \frac{\langle \sigma v (\DN[k]+\DN[2] \rightarrow \DN[k+2])\rangle}{\sigma_\circ\langle v \trans \rangle_\circ} \right)\,.
\end{align}
Working in the saturation limit, and taking the large $k$ continuum limit, the cross sections will scale as $\langle \sigma v (\DN[k]+\DN[2] \rightarrow \DN[k+2])\rangle \simeq 
\sigma_\circ \langle v \trans \rangle_\circ  k^{\frac{2}{3}}$, and we have

\begin{align}
  \frac{\partial}{\partial\gamma}\left(k^{\frac{2}{3}}y_k \right)= - \xi \frac{\partial}{\partial k^{\frac{1}{3}}} \left( k^{\frac{2}{3}}y_k\right)\,,
\label{eq:bottleneck_wave}
\end{align}
where $\xi = 2y_2 /3$ is approximately constant. 
Analytically, \Eq{eq:bottleneck_wave} is a simple linear wave equation describing a distribution moving forward in $k^{\frac{1}{3}}$-space with speed $\xi$. This implies that, for nucleation site populations with roughly the same size $k$, the distribution will remain peaked in $k^{\frac{1}{3}}$-space at later times. Inclusions of large-large interactions will change the shape of the distributions, but should remain subdominant as long as the nucleation sites remain sparse. The average nugget size for the nucleation sites exiting synthesis can then be approximated by
\begin{align}
  \bar k_{\rm fo} \sim \left(\xi \gamma_{\rm max}\right)^3
\end{align}
with $\gamma_\text{max}$ given by \Eq{gammaparam_bottleneck}. Explicitly, 
\begin{multline}
\gamma_\text{max} \approx {4\times 10^{4}} \sqrt{g_{* S}^\text{syn} \over g_*^\text{syn}} \sqrt{g_{* S}^\text{syn} \over 10 } {\Tgambbd \over  \TXbbd} \left(\TXbbd \over \BE_2 / 28 \right)^{2} \\
\times  \left( r_0 m_X \over 23 \right)^2 \left(10 \text{GeV} \over m_X \right)^2 \left(400 \BE_2 \over  m_X \right)^{2} \left( m_X \over 10 \bar  m _X \right)^{2},
 \label{eq: gamma 3}
\end{multline}
where fiducial values correspond roughly to the benchmark parameters $\alpha_\phi = 0.1$, ${m_X = 10\, \text{GeV}}$, ${m_\phi = 10\, \text{MeV}}$. 

The wave equation description breaks down when $\DN[2]$ starts to be depleted, which happens when the fractional DM number contained in the nucleation sites become $\mathcal{O}(1)$. If we assume that the probability of a single $\DN[2]$ squeezing through the bottleneck at the beginning of synthesis is $2p$, then the number density of the nucleation sites will be $p n_X$. The nucleation sites will evolve linearly until the fractional DM number in nucleation sites, $p (\xi \gamma)^3 $, is roughly $1/2$, at the transition interaction time
\begin{align}
  \gamma_\text{trans} \gtrsim \frac{1}{\xi}\frac{1}{\sqrt[3]{2p}}\,.
\end{align}
At this point, the average nucleation size will be $\bar k \sim 1/(2p)$. Beyond this point, the $\DN[2]$ distribution is expected to rapidly become depleted, and the subdominant large-large fusion will become significant. If $\gamma_\text{trans}^3 < \gamma_\text{max}^{6/5}$, the discussion in Sec.~\ref{eq:nugget_formation_noBN} will then apply once again, with a scaling law $\bar k \sim \gamma^{5/6}$.

\begin{figure}
\begin{minipage}{0.5\textwidth}
\includegraphics[width=1\textwidth]{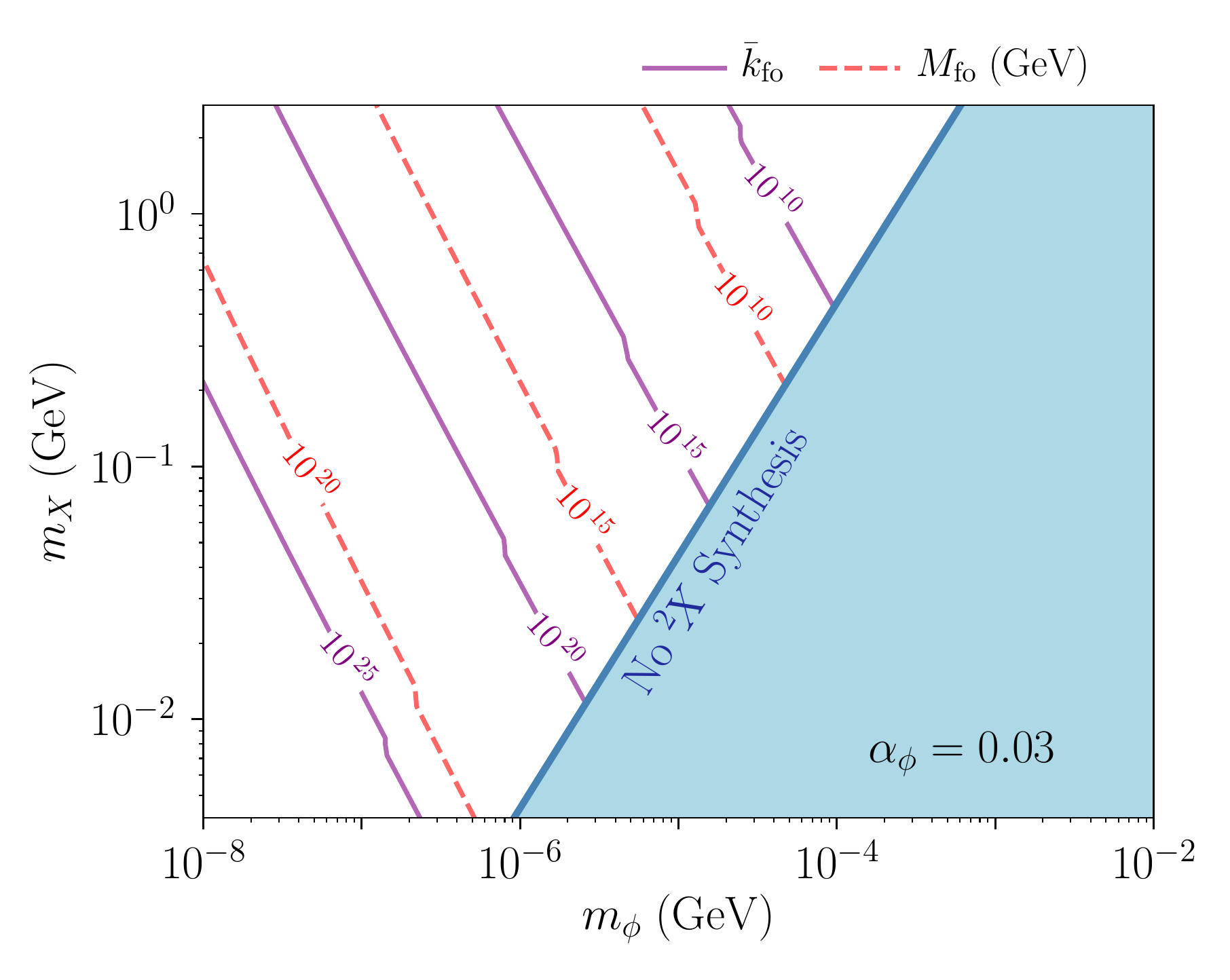}~
\end{minipage}
\begin{minipage}{0.48\textwidth}
\includegraphics[width=1\textwidth]{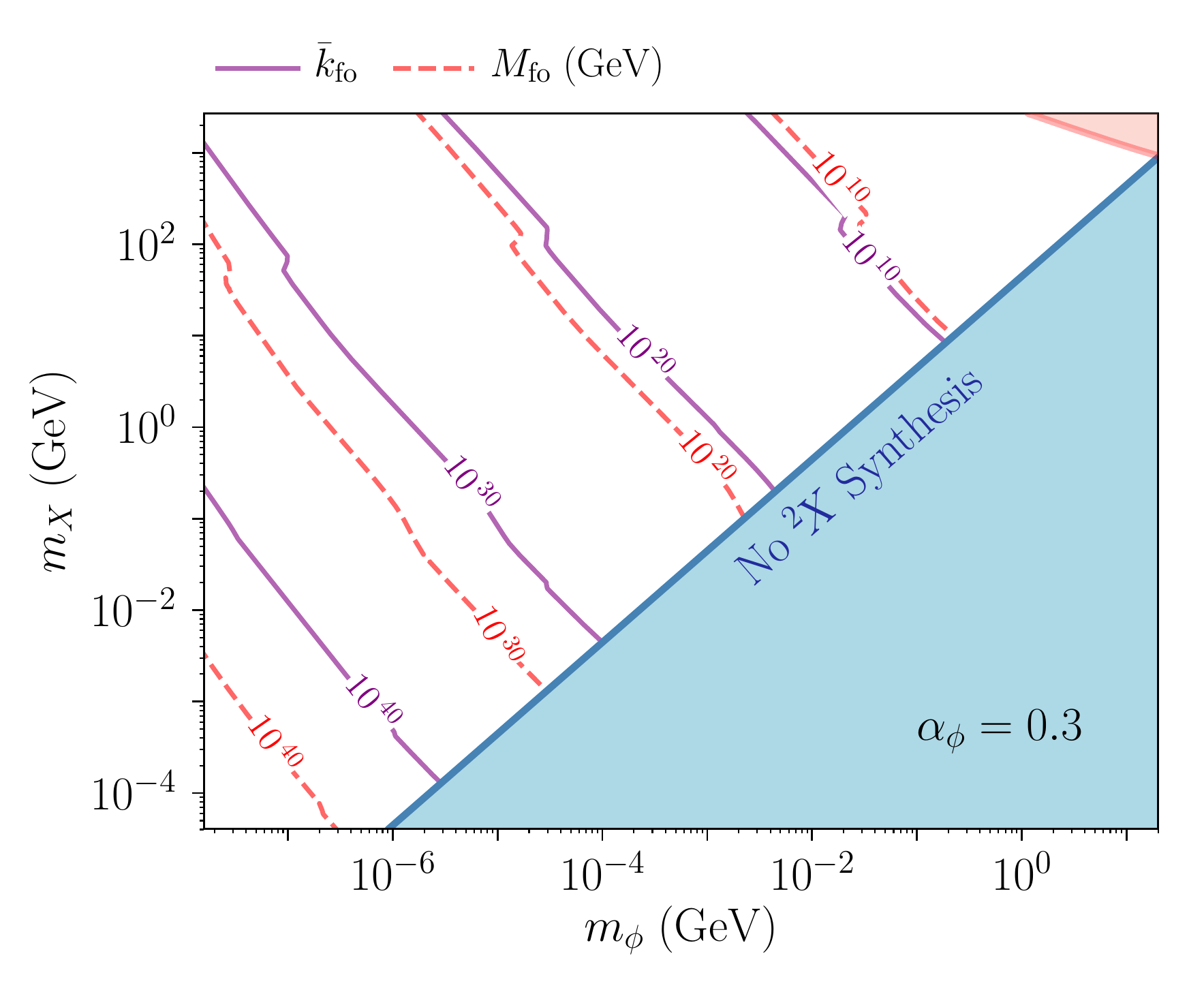}
\end{minipage}
\caption{Contours of the nugget number, $k_\text{fo}$, (\emph{dashed red}), and the typical mass of the nuggets, $M_{\text{fo}}$ (\emph{solid purple}) for the large nugget populations exiting darkleosynthesis when a significant bottleneck is present. The coupling is fixed at $\alpha_\phi=0.03$ (left) and $\alpha_\phi=0.3$ (right). Similar to Fig.~\ref{fig: Nmax}, the blue shaded region corresponds to inefficient $\DN[2]$ fusion due to small binding energy, and the upper (lower) cutoff for $m_X$ corresponds to $\DN[2]$ fusion rate always smaller than Hubble and $T_\text{synth} \lesssim 10 T_\text{eq}$ respectively.}\label{fig: bottleneck plot}
\end{figure}

Fig.~\ref{fig: bottleneck plot} shows the nugget sizes and masses for the nucleation sites exiting synthesis. We have assumed that $p$ is small enough that the $\DN[2]$ population always dominates. Compared to Fig.~\ref{fig: Nmax}, the final nugget number and masses are significantly larger due to $\gamma_{\rm max}^3$ scaling. This is expected as the fusion rate is controlled by the $\DN[2]$ densities which remain relatively large. One may be concerned that the local $\DN[2]$ density within a Hubble volume of a nugget may be depleted before $\gamma_{\rm max}$ is reached, which would render our calculation invalid. Such a requirement will impose a lower bound on $p \gtrsim H^3/n_X \sim {\bar m_X \over  \text{GeV}} {T^3 \over \eta m_\text{Pl}^3}$, which is negligibly small when the temperature is of order GeV or smaller.

\subsection{Nugget Distribution and Bottlenecks}
\label{sec:nuggetmassfunction_bottleneck}

Here we present a numerical investigation of the effects of bottlenecks on the nugget distribution. In Sec.~\ref{sec:bottleneck} we argued that a significant bottleneck might occur due to $\DN[3]$ \emph{and} $\DN[4]$ both being unstable, while isolated unstable nuggets do not cause significant changes in the nugget distributions.

\begin{figure}
\begin{minipage}{.45\textwidth}
  \includegraphics[trim={0cm 0cm 0cm 0cm},clip,width=1.0\linewidth]{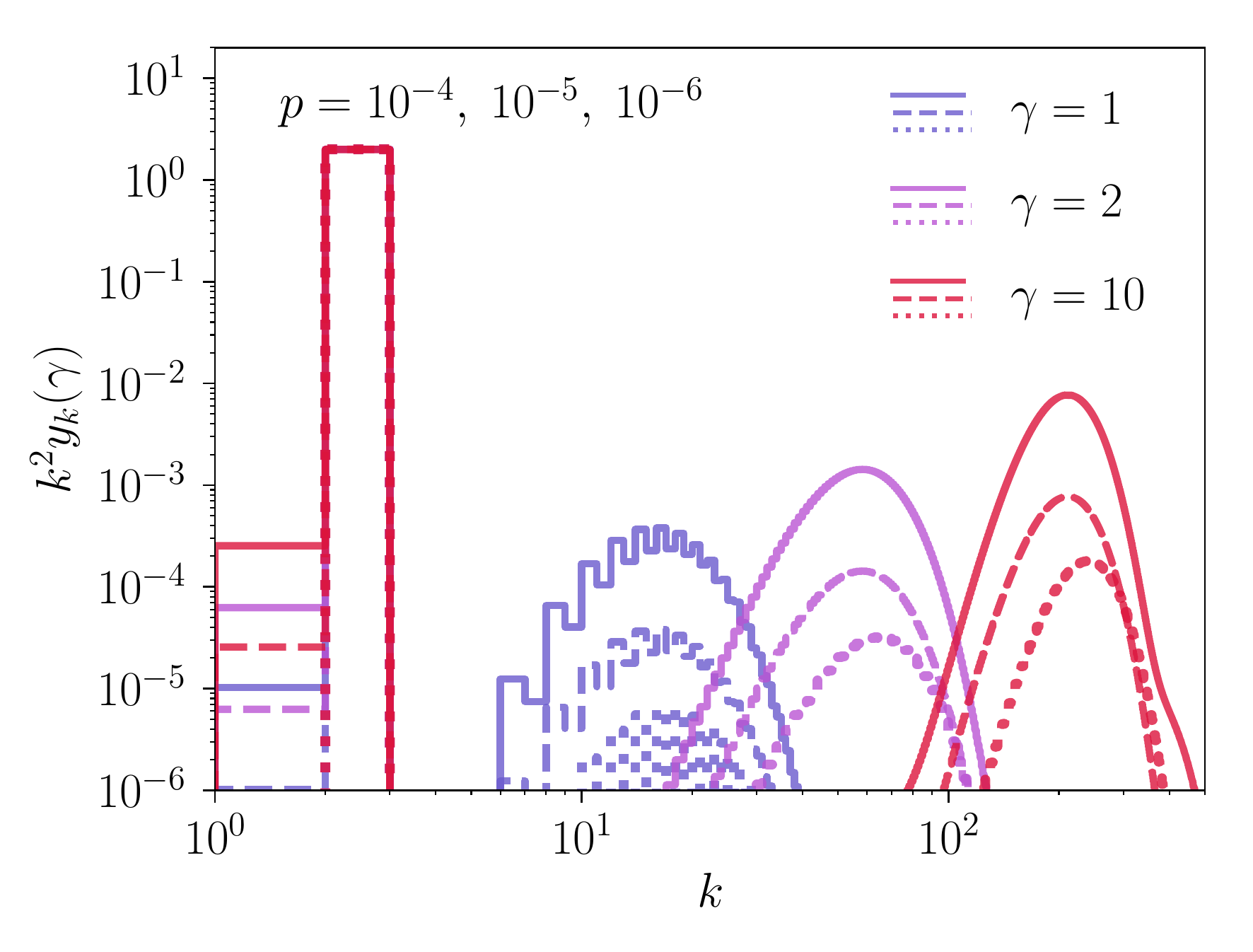}
\end{minipage}
\hspace{.3cm}
\begin{minipage}{.45\textwidth}
\includegraphics[trim={0cm 0cm 0cm 0cm},clip,width=1.0\linewidth]{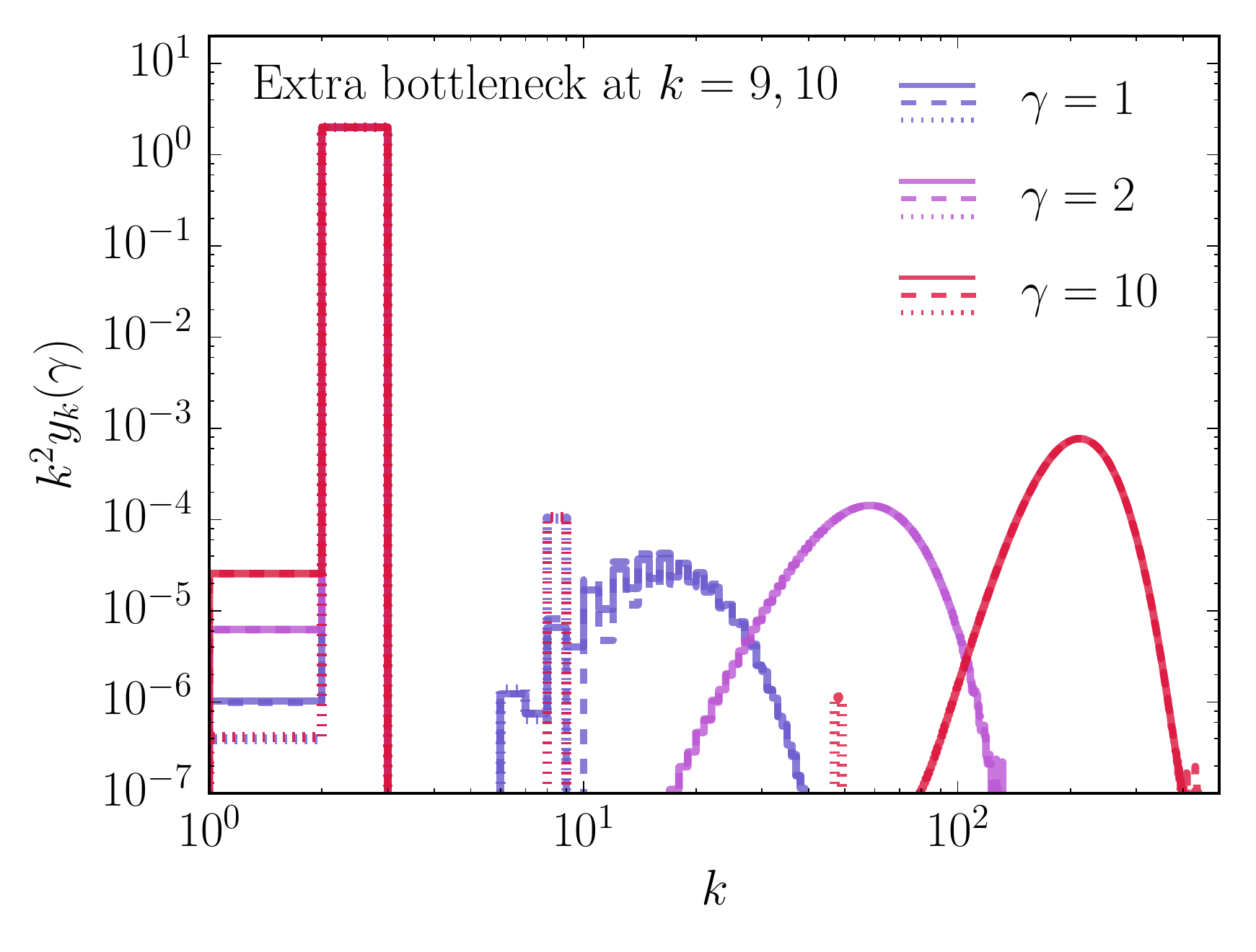}
\end{minipage}
\caption{Left: Nugget mass function at different interaction time $\gamma$, with two bottlenecks at $k=3,4$. The solid/dashed/dotted corresponds to $p$ equals to $10^{-4}/10^{-5}/10^{-6}$ respectively. Right: Nugget mass function for fixed $p=10^{-5}$, with solid/dashed/dotted corresponding no additional bottleneck, additional bottleneck at $k=9$, and additional bottleneck at $k=9,10$. The solid and dotted lines are indistinguishable at $\gamma=2,10$, while for the dotted line only a very small population extends beyond the $k=8$ bin. }
\label{fig:synth_bottleneck_p}
\end{figure}

In order to simulate nugget evolution in this situation, we follow the same setup in Sec.~\ref{sec:nuggetmassfunction}, and use the cross sections in \Eq{eq: kernel}. Though \Eq{eq: kernel} is modeled on large-large fusion, the velocity dependence approaches a constant for small-large fusions and describes the bottleneck scenario well.
 The large-large fusion cross-sections is not fully captured by the kernel, as its temperature dependence is different form the small-large case. Although contributions from large-large interactions is sub-dominant whenever $p \ll  \sqrt{\Tbbd \bar m_X / m_X^2} \sim \alpha_\phi^{7/8} (m_\phi/m_X)^{1/4} / 10$, which is relevant for the benchmark cases we have. 
The Boltzmann equation is truncated at $k\le 1000$, such that all branchings into higher $k$ nuggets are fixed to zero. We also utilize the energy scaling branching ratios, though different branching assumptions do not change the quantitative behavior significantly. The initial conditions are chosen such that $2y_2=1-p$ and $6 y_6=p$. To simulate the effects of bottlenecks, we fix $y_{k}=0$ and $\Gamma_{kl}=0$ if $\DN[k]$ or $\DN[l]$ is unstable. Left of Fig.~\ref{fig:synth_bottleneck_p} shows the mass distributions for different choices of $p$.  We see that in general, the nugget distributions are separated into two distinct populations: the small $\DN[2]$ population and large nugget nucleation sites. The distributions for the nucleation sites are strongly peaked, and move forward rapidly as $\gamma$ increases. The total fraction of DM in the nucleation sites increases as well. Variations in $p$ simply modify the total nucleation site density. At large $\gamma$, one expects the $\DN[2]$ density to appreciably decrease eventually, although a numerical confirmation is impractical, as it requires including exponentially more terms in the Boltzmann equations. 

The right of Fig.~\ref{fig:synth_bottleneck_p} also shows a comparison of different nugget functions when including extra bottlenecks at $k=9$ for the dashed line (and also $10$ for the dotted line), while fixing $p=10^{-5}$. With only one extra bottleneck at $k=9$, the mass functions quickly move beyond the bottleneck and become indistinguishable from the ones without the $k=9$ bottleneck. While for the case with two bottlenecks at $k=9,10$, no larger nuggets are produced beyond $k=8$. This confirms our expectations, where isolated bottlenecks do not change the nugget distribution while multiple consecutive bottlenecks can bring fusion processes to a halt.

\section{Conclusions}

We have studied the early Universe synthesis of many-particle bound states of ADM, or nuggets.  We unified the treatment of synthesis with a quantitative calculation, in our companion paper \cite{Gresham:2017zqi}, of the properties of the bound state, utilizing effective field theory tools developed in nuclear physics.  Within our model, the typical nugget size exiting early Universe synthesis can be many orders of magnitude larger than has been estimated in previous treatments of many-particle bound states of dark matter. 
 We derived a nugget mass function, describing the dark matter energy distribution in nuggets of different sizes, from the Boltzmann equation.  We found that inclusion of the analog of baryon-number-transfer nuclear fusion reactions in addition to radiative fusion reactions generates a broader nugget distribution. 

With a quantitatively derived nugget mass function exiting early Universe synthesis, we are now in a position to study the late Universe cosmology of ADM nuggets.  This is the subject of our next investigation.

\section*{Acknowledgments}
We thank Dorota Grabowska, Tom Melia, Mark Wise, and Yue Zhang for comments on the draft and Martin Savage for discussions. HL and KZ are supported by the DoE under contract DE-AC02-05CH11231. HL is also supported by NSF grant PHY-1316783. MG is supported by the Murdock Charitable Trust. 

\bibliography{ADMNuggetsMerged}

\end{document}